\documentclass[prb,reprint,amsmath,amssymb,showpacs,twocolumn]{revtex4-1}

\usepackage[dvipdfmx]{graphicx}
\usepackage{bm}
\usepackage{mathrsfs}
\usepackage{color}

\DeclareMathOperator{\diag}{diag}

\begin{document}
\title{Topologically stable gapless phases in nonsymmorphic superconductors}
\author{Shingo Kobayashi$^{1,2}$, Youichi Yanase$^3$, and Masatoshi Sato$^4$}
\affiliation{$^1$Department of Applied Physics, Nagoya University, Nagoya 464-8603, Japan \\ 
		$^2$Institute for Advanced Research, Nagoya University, Nagoya 464-8601, Japan \\
		$^3$Department of Physics, Kyoto University, Kyoto 606-8502, Japan \\
		$^4$Yukawa Institute for Theoretical Physics, Kyoto University, Kyoto 606-8502, Japan}

\date{\today}

\begin{abstract}
We study topological stability of nodes in nonsymmorphic superconductors (SCs). In particular, we demonstrate that line nodes in nonsymmirphic odd-parity SCs are protected by the interplay between topology and nonsymmorphic symmetry. As an example, it is shown that the $E_{2u}$-superconducting state of UPt$_3$ hosts the topologically stable line node at the Brillouin zone face. Our theory indicates that the existence of spin-orbit coupling is essential for protecting such a line node, complementing the Norman's group theory argument.  
Developing the topological arguments, we also argue generalization to point nodes and to other symmetry cases beyond the group theory arguments. 
\end{abstract}
\maketitle
\section{Introduction}
Nontrivial node structures are a salient feature in the unconventional superconductors (SCs), offering valuable clues to understanding of symmetry of Cooper pairs. The node structures are detected by the power law behaviors of temperature dependence such as the specific heat and NMR relaxation rates~\cite{Legget:1975,Sigrist:1991}. From the 1980s,  superconductivity in heavy fermion materials has attracted much attention as candidates of unconventional SCs~\cite{Pfleiderer:2009}. At early stage, the group theory is applied to the superconducting states in order to limit possible Cooper pairs~\cite{Anderson:1984} by use of spin-orbit coupling (SOC) and crystal structure in heavy fermion materials. From the group theoretical ground, Blount showed the impossibility of line nodes in odd-parity SCs in the presence of SOC~\cite{Blount:1985}, which is called the Blount's theorem. To the contrary, real candidate materials of heavy fermion odd-parity SCs such as UPt$_3$ have often suggested the existence of line nodes~\cite{Joynt:2002}. To resolve the contradiction, Norman pointed out the possibility of a stable line node on the Brillouin zone (BZ) face in odd-parity SCs due to nonsymmorphic symmetry~\cite{Norman:1995}, which is a counter example of the Blount's theorem. Furthermore, Mickliz and Norman proved that two-fold screw symmetry generally forces an odd-representation of Cooper pair to vanish on the BZ face perpendicular to the screw axis within the group theory~\cite{Micklitz:2009}.

Recently, unconventional SCs have been seen in a new light, i.e., from the viewpoint of topology~\cite{Volovik:2003,Sato:2006,Beri:2010,Horava:2005,Zhao:2013v1,Shiozaki:2014,Kobayashi:2014,Chiu:2014,SAYang:2014,Chiu:2015}. Stability of nodes is given by topological numbers. The topological approach enables us to classify symmetric and accidental nodes in a unified way and may connect topological structures of bulk nodes to surface zero-energy states via the bulk-edge correspondence~\cite{Tanaka:2010,Sato:2011,Yada:2011,Tanaka:2012,Schnyder:2011,Schnyder:2012,Brydon:2011,Matsuura:2013,Schnyder:2015,Kobayashi:2015}. Using this method, two of the present authors proved the topological version of the Blount's theorem~\cite{Kobayashi:2014} and updated the conventional Blount's theorem by connecting a bulk line node with a surface zero-energy flat band instability. At the same time, the reinterpretation may also raise new theoretical questions regarding the connection between the Norman's argument and the topological approach, i.e, the topological stability of line nodes in nonsymmorphic odd-parity SCs. 

Besides unconventional SCs, similar gapless phases have been argued in the context of topological semimetals, such as Dirac/Weyl semimetals~\cite{Murakami:2007,Young:2012,Wang:2012,Wang:2013,Steinberg:2014,Yang:2014,Xu:2014,Liu1:2014,Neupane:2014,Liu2:2014,Jeon:2014,Yi:2014,Borisenko:2014,Wan:2011,Burkov:2011,Weng:2015v1} and line nodal semimetals~\cite{YKim:2015,Yu:2015,Phillips:2014,Mullen:2015,Xie:2015, Weng:2015v2,Chen:2015v1,Yamakage:2016,Bian:2016v1,Bian:2016v2}. Among them, tight-binding model studies in orthorhombic perovskite SrIrO$_3$ showed a stable line nodes at the BZ face~\cite{Carter:2012,Chen:2015v2,HSKim:2015}. Because of nonsymmorphic symmetry and strong SOC in SrIrO$_3$, the line node is topologically protected~\cite{Fang:2015v1}. Recently, versatile topological semimetals~\cite{Watanabe:2015,SMYoung:2015,Watanabe:2016,Liang:2016,Wieder:2016,BJYang:2016,XYZhan:2016} and insulators~\cite{CXLiu:2014,Fang:2015v2,Shiozaki:2015,ZWang:2016,Alexandradinata:2016,Shiozaki:2016,QZWang:2016v1,Mong:2010,Dong:2016,QZWang:2016v2,Varjas:2015,Sahoo:2015,LLu:2016,PYChang:2016} with nontrivial influence of nonsymmorphic symmetry have been anticipated theoretically. 

  In this paper, we establish a general theory to treat topological stability of nodes in nonsymmorphic SCs. Our theory enables us to take into account nonsymmorphic crystal in the topological manner and is natural extension of the previous work~\cite{Kobayashi:2014}. The obtained results include the topological Blount's theorem. In a generalized framework, we will find that the line node proposed by Norman is exactly protected by interplay between topology and nonsymmorphic symmetry, the stability of which is characterized by a mirror topological number. Besides the topological number, we also reveals that SOC plays a central role in protecting the line node. In the absence of SOC, the Fermi surface acquires a four-fold degeneracy at the BZ face, and the line node disappears. We apply our theory to the $E_{2u}$ superconducting state of UPt$_3$ and show the existence of nonsymmorphic-symmetry-protected nodal rings at the BZ face, by taking into account an anti-symmetric SOC.  In addition, the topological approach predicts nontrivial nonsymmrphic symmetry protected nodes beyond the Norman's argument. Thus, our results not only connect the group theory studies with the topological classification, but also provide a new guiding principle in searching for nonsymmorphic symmetry protected nodes.  
  
  The paper is organized as follows. In Sec.~\ref{sec:fomulation}, we construct the Bogoliubov-de Gennes (BdG) Hamiltonian, with taking into account nonsymmrphic crystals. This part is at the heart of the mechanism of nonsymmorphic symmetry protected nodes. In Sec.~\ref{sec:line-node}, stability of line nodes in nonsymmorphic odd-parity SCs is discussed in two different ways: group theoretical classification of possible Cooper pairs in Sec.~\ref{subsec:Group} and topological classification of BdG Hamiltonians in  Sec.~\ref{subsec:Topology}. We apply the topological argument to the $E_{2u}$-representation superconducting state of UPt$_3$ in Sec.~\ref{subsec:UPt3}. In Sec.~\ref{sec:general}, we mention possible generalization of nonsymmorphic symmetry protected nodes. Finally, we summarize this paper in Sec.~\ref{sec:summary}.

\section{Formulation}
\label{sec:fomulation}
First, we generalize the basis function of the underlying Hamiltonian in order to take into account nonsymmorphic crystals~\cite{footnote1}. A nonsymmorphic crystal has at least two atoms in the unit cell, and these atoms are separated by a non-primitive lattice vector. To involve the non-primitive lattice vector in a tight-binding Hamiltonian, we use L\"{o}wdin orbitals $\varphi_{ \alpha} (\bm{r}- \bm{R} - \bm{r}_{\alpha})$~\cite{Lowdin:1950}, where $\bm{R}$ is a Bravais lattice (BL) vector and $\bm{r}_{\alpha}$ denotes a position of an atom $\alpha$. Here, $\alpha$ ($\alpha=1,\cdots,m$) describes spin, sublattice indices, and orbitral degrees of freedom. The wave function centered at different sites (or with different indices $\alpha$) are orthogonal to each other. The basis function, which has a discrete translational invariance in terms of BL vectors, is given by the linear combination of L\"{o}wdin orbitals: 
\begin{align}
\phi_{\bm{k}, \alpha} (\bm{r}) = \frac{1}{\sqrt{N}} \sum_{\bm{R}} e^{i \bm{k} \cdot (\bm{R}+\bm{r}_{\alpha})} \varphi_{ \alpha} (\bm{r}- \bm{R} - \bm{r}_{\alpha}),
\end{align} 
where $N$ is the number of primitive unit cells in the crystal. The function $\phi_{\bm{k},\alpha}$ obeys the Bloch condition: $\phi_{\bm{k},\alpha} (\bm{r}+\bm{R}) = e^{i \bm{k}\cdot \bm{R}} \phi_{\bm{k},\alpha} (\bm{r})$ and, due to a non-primitive lattice vector $\bm{r}_{\alpha}$, it satisfies the additional condition:
$ \phi_{\bm{k+G} , \alpha} (\bm{r}) = e^{i \bm{G} \cdot \bm{r}_{\alpha}}  \phi_{\bm{k} , \alpha}  (\bm{r})$,
where $\bm{G}$ is a reciprocal lattice (RL) vector. If $\bm{r}_{\alpha} = \bm{0}$, the L\"{o}wdin orbital reduces to the Wanner function.  Using the L\"{o}wdin orbitals,  the tight-binding Hamiltonian is given by~\cite{Alexandradinata:2016}
\begin{align}
H_{\alpha\beta}(\bm{k}) = \int d \bm{r} \; \phi^{\ast}_{\bm{k},\alpha}(\bm{r})  \mathcal{H}  \phi_{\bm{k},\beta}(\bm{r}), \label{eq:TBH}
\end{align} 
where $\mathcal{H}$ is the single-particle Hamiltonian. The tight-binding Hamiltonian satisfies
\begin{align}
 H_{\alpha \beta} (\bm{k}+\bm{G}) = e^{- i \bm{G} \cdot \bm{r}_{\alpha}}H_{\alpha \beta} (\bm{k}) e^{i \bm{G} \cdot \bm{r}_{\beta}}. \label{eq:TBH_cond}
\end{align}
 
 We phenomenalogically model nonsymmorphic superconductors (SCs) using the L\"{o}wdin orbitals. We introduce a creation operator of the wave function $\phi_{\bm{k},\alpha}$,~\cite{Alexandradinata:2016}
 \begin{align}
  c_{\bm{k},\alpha}^{\dagger} = \frac{1}{\sqrt{N}} \sum_{\bm{R}} e^{i\bm{k} \cdot (\bm{R}+\bm{r}_{\alpha})} c_{\alpha} (\bm{R}+\bm{r}_{\alpha})^{\dagger}, \label{eq:Lowdin-operator}
 \end{align} 
where $c_{\alpha} (\bm{R}+\bm{r}_{\alpha})^{\dagger}$ is a creation operator of electron with index $\alpha$ located at $\bm{R}+\bm{r}_{\alpha}$. Equation~(\ref{eq:Lowdin-operator}) satisfies $ c_{\bm{k}+\bm{G},\alpha}^{\dagger} = e^{i \bm{G}\cdot \bm{r}_{\alpha}} c_{\bm{k},\alpha}^{\dagger}$. The Bogoliubov-de Gennes (BdG) Hamiltonian is given by
 \begin{align}
  H_{\rm BdG} = \frac{1}{2}\sum_{\bm{k},\alpha,\beta} \left( c_{\bm{k},\alpha}^{\dagger}, c_{-\bm{k},\alpha} \right) \tilde{H} (\bm{k}) \left( \begin{array}{@{\,} c @{\,}} c_{\bm{k},\beta} \\ c_{-\bm{k}, \beta}^{\dagger}\end{array} \right),  \label{eq:BdG1}
 \end{align} 
 with
 \begin{align}
  \tilde{H} (\bm{k})= \begin{pmatrix} \mathcal{E}_{\alpha \beta} (\bm{k})  & \Delta_{\alpha \beta} (\bm{k})\\ \Delta_{\alpha \beta}(\bm{k})^{\dagger}  & - \mathcal{E}_{\alpha \beta} (-\bm{k})^T\end{pmatrix} ,\label{eq:BdG2}
 \end{align} 
 where $\mathcal{E}_{\alpha \beta}(\bm{k}) = H_{\alpha \beta} (\bm{k}) -\mu \delta_{\alpha \beta}$ is the normal Hamiltonian, the gap function $\Delta_{\alpha \beta} (\bm{k})$ satisfies $\Delta_{\alpha \beta} (-\bm{k}) = -\Delta_{\beta \alpha} (\bm{k})$ due to the Fermi statistics, and $\mu$ is the chemical potential. Since the gap function should be consistent with the structure of nonsymmorphic crystals, we requires $\Delta_{\alpha \beta} (\bm{k}+\bm{G}) =  e^{- i \bm{G} \cdot \bm{r}_{\alpha}} \Delta_{\alpha \beta} (\bm{k})  e^{ i \bm{G} \cdot \bm{r}_{\beta}}$. Combining it with Eq.~(\ref{eq:TBH_cond}), the BdG Hamiltonian has the constraint under an RL vector $\bm{G}$: 
 \begin{align}
  \tilde{H}(\bm{k} + \bm{G}) = \tilde{V}_{\bm{G}} \tilde{H}(\bm{k}) \tilde{V}_{\bm{G}}^{\dagger}, \ \  \tilde{V}_{\bm{G}} = \begin{pmatrix} V_{\bm{G}} & 0 \\ 0 & V_{\bm{G}} \end{pmatrix}, \label{eq:plusG}
 \end{align}
 with $V_{\bm{G}} = \diag \left[ e^{-i \bm{G} \cdot \bm{r}_1}, \cdots,  e^{-i \bm{G} \cdot \bm{r}_m} \right]$ ($\alpha = 1,\cdots,m$). 
 
 In the following, we summarize discrete symmetries that are relevant to stability of nodes. To start with, we introduce particle-hole symmetry (PHS), time-reversal symmetry (TRS), and spatial-inversion symmetry (IS) as follows.
\begin{align}
{\rm PHS:} \ \ C\tilde{H}(\bm{k})C^{\dagger} = - \tilde{H}(-\bm{k}), \ \ C = \begin{pmatrix} 0 & \delta_{\alpha \beta} \\\delta_{\alpha \beta} & 0\end{pmatrix} K, \label{eq:PHS}
\end{align} 
\begin{align}
{\rm TRS:} \ \ \ \mathcal{T} \tilde{H}(\bm{k}) \mathcal{T}^{\dagger} = \tilde{H} (-\bm{k}), \ \ \mathcal{T} = \begin{pmatrix} T_{\alpha \beta} & 0 \\ 0 & T^{\ast}_{\alpha \beta} \end{pmatrix}, \ \ \label{eq:TRS}
\end{align}

  \begin{align}
{\rm IS:}  \ \ \ \mathcal{P} \tilde{H}(\bm{k}) \mathcal{P}^{\dagger} = \tilde{H} (-\bm{k}), \ \  \mathcal{P} = \begin{pmatrix} P_{\alpha \beta} & 0 \\ 0 & \eta_P P^{\ast}_{\alpha \beta} \end{pmatrix}. \label{eq:IS}
   \end{align}
Here, $T \equiv U_t K$ and $U_t$ and $P$ are $m \times m$ unitary matrices satisfying $U_t = -U_t^{T}$ and  $P^2 =\bm{1}_{m}$. $K$ is the complex conjugation operator and $\bm{1}_m$ is the identity matrix with rank $m$. From Eqs.~(\ref{eq:TRS}) and (\ref{eq:IS}), TRS and IS, respectively, require $ U_t H(\bm{k}) U_t^{\dagger} = H^{\ast}(-\bm{k})$ and $U_t \Delta(\bm{k}) U_t^{T} = \Delta^{\ast} (-\bm{k}) $, and $ P H(\bm{k}) P^{\dagger} = H(-\bm{k})$ and $P \Delta(\bm{k}) P^{T} = \eta_P \Delta (-\bm{k})$, where $\eta_P$ describes the parity of the gap function, i.e., $\eta_P = +1$ for even parity and $\eta_P =-1$ for odd parity. For even parity gap functions, we have $[C,\mathcal{P}]=0$, while for odd parity gap functions, $\{C,\mathcal{P}\}=0$. Hereafter, we assume $[C,\mathcal{T}]=[\mathcal{P},\mathcal{T}]=0$ unless otherwise specified.
 
 In addition to the non-spatial symmetries, crystal symmetry may stabilize nodal structure. An element of a space group $G$ is given as $\{g | \bm{\tau}\}$ with a point group element $g$ and a translation $\bm{\tau}$. Under $\{g | \bm{\tau}\}$, $\bm{x}$ transforms as $\bm{x} \to D(g) \bm{x}+\bm{\tau}$. For $\{g | \bm{\tau}\}$, $c_{\bm{k} \alpha}^{\dagger}$ transforms as (see Appendix~\ref{app:action})
 \begin{align}
  \{ g|\bm{\tau} \} c_{\bm{k},\alpha}^{\dagger} \{ g|\bm{\tau}\}^{-1} &= e^{- iD(g) \bm{k} \cdot \bm{\tau}} c_{D(g)\bm{k},\beta}^{\dagger} U_{\beta \alpha} (g) \notag \\
  &\equiv c^{\dagger}_{D(g) \bm{k},\beta} \, D_{\bm{k}}(\{g|\bm{\tau}\})_{\beta \alpha}, \label{eq:GA}
 \end{align}
 where $D(g)$ and $U(g)$ are matrix representations of $g$ in real space and space of $\alpha$, respectively. When $\{ g_1 | \bm{\tau}_1\}$ and $\{ g_2 | \bm{\tau}_2\}$ are elements of the little group leaving $\bm{k}$ invariant, the associative property of $D_{\bm{k}}$, $D_{\bm{k}} (\{g_1|\bm{\tau}_1\})D_{\bm{k}} (\{g_2|\bm{\tau}_2\})=D_{\bm{k}} (\{g_1g_2|D(g_1)\bm{\tau}_2+\bm{\tau}_1\})$, leads to  
\begin{align}
 U(g_1) U(g_2) = \omega_{g_1,g_2}^{\bm{k}}U(g_1g_2), \label{eq:asso1}
\end{align} 
 where $\omega_{g_1,g_2}^{\bm{k}} \equiv e^{i\bm{k}\cdot (D(g_1)^{-1} \bm{\tau}_1 -D(g_2)^{-1} D(g_1)^{-1}\bm{\tau}_1) }$ is a factor system in the group theory~\cite{Bradley:1972}. Here, the factor system $\omega_{g_1,g_2}^{\bm{k}}$ is nontrivial only if $\bm{k}$ is in a high-symmetric subspace on the BZ face.
Furthermore, if $U(g_1g_2)=t_{g_1,g_2} U(g_2g_1)$ with $t_{g_1,g_2} = \pm 1$, the commutation relation between $D_{\bm{k}}(\{g_1| \bm{\tau}_1\})$ and $D_{\bm{k}}(\{g_2| \bm{\tau}_2\})$ becomes
\begin{align}
 U(g_1)U(g_2) = t_{g_1,g_2} \alpha_{g_1,g_2}^{\bm{k}} U(g_2) U(g_1), \label{eq:asso2} 
\end{align} 
where $\alpha_{g_1,g_2}^{\bm{k}} \equiv \omega_{g_1,g_2}^{\bm{k}}/\omega_{g_2,g_1}^{\bm{k}}$.

From Eq.~(\ref{eq:GA}), $H_{\rm BdG}$ in Eq.~(\ref{eq:BdG1}) transforms as
\begin{align}
 H_{\rm BdG} \to &\sum \bm{c}^{\dagger}_{D(g)\bm{k} } D_{\bm{k}} (\{g|\bm{\tau}\}) \mathcal{E} (\bm{k}) D_{\bm{k}}(\{g|\bm{\tau}\})^{\dagger}  \bm{c}_{D(g)\bm{k}} \notag \\
   &+ \sum \bm{c}^{\dagger}_{D(g)\bm{k}} D_{\bm{k}} (\{g|\bm{\tau}\}) \Delta (\bm{k}) D_{-\bm{k}} (\{g|\bm{\tau}\})^T \bm{c}_{-D(g)\bm{k}} \notag \\ 
   &+ \cdots \label{eq:BdG-A}
\end{align}
under $\{g |\bm{\tau}\}$. Since the normal Hamiltonian is invariant under $G$, we have
\begin{align}
D_{\bm{k}} (\{g|\bm{\tau}\}) \mathcal{E} (\bm{k}) D_{\bm{k}} (\{g|\bm{\tau}\})^{\dagger} = \mathcal{E} (D(g)\bm{k}).  \label{eq:BdG-B}
\end{align}
Moreover, in order for $\{g |\bm{\tau}\}$ to be symmetry of the superconducting state, the gap function should obey
\begin{align}
D_{\bm{k}} (\{g|\bm{\tau}\}) \Delta(\bm{k}) D_{-\bm{k}} (\{g|\bm{\tau}\})^T = \eta_{C,g} \Delta (D(g)\bm{k}), \label{eq:BdG-C}
\end{align}
with $\eta_{C,g} = \pm 1$. For $\eta_{C,g} = 1 (-1)$, the right hand side of Eq.(\ref{eq:BdG-A}) coincides with $H_{\rm BdG}$ trivially (by performing the $\pi$-gauge rotation of $c_{\bm{k} \alpha}^{\dagger}$). The phase factors  $e^{- iD(g) \bm{k} \cdot \bm{\tau}}$ are canceled in Eqs.~(\ref{eq:BdG-B}) and (\ref{eq:BdG-C}), so we have
\begin{align}
&U(g)\mathcal{E} (\bm{k}) U (g)^{\dagger}= \mathcal{E} (D(g)\bm{k}), \label{eq:BdG-D}\\
&U(g)\Delta (\bm{k}) U(g)^{T} = \eta_{C,g} \Delta (D(g)\bm{k}). \label{eq:BdG-E}
\end{align}
  In the matrix form of the BdG Hamiltonian, Eqs.~(\ref{eq:BdG-D}) and (\ref{eq:BdG-E}) are summarized as 
\begin{align}
 \tilde{U}(g) \tilde{H} (\bm{k}) \tilde{U}(g)^{\dagger}  = \tilde{H} (D(g) \bm{k}), \label{eq:UHU}
\end{align}
 with $\tilde{U}(g) = \diag [U(g), \eta_{C,g} U(g)^{\ast}]$. We also have $C\tilde{U}(g) =\eta_{C,g} \tilde{U}(g) C$.
 
 Since we are interested in the influence of the crystal symmetry on nodes, we focus on the behavior of the BdG Hamiltonian near a node at $\bm{k}_0$, where the position of a node is defined by $\det[ \tilde{H} (\bm{k}_0)] =0$. We assume that $\bm{k}_0$ lies in a high symmetry subspace of BZ and $\{g|\bm{\tau}\}$ belong to the little group of $\bm{k}_0$, i.e., $D(g)\bm{k}_0 -\bm{k}_0$ is an RL vector. With the condition~(\ref{eq:plusG}), the space group operation $\{g|\bm{\tau}\}$ on the BdG Hamiltonian at $\bm{k} + \bm{k}_0$ yields
 \begin{align}
  &\tilde{U}_{\bm{k}_0}(g) \tilde{H}(\bm{k} +\bm{k}_0) \tilde{U}_{\bm{k}_0}(g)^{\dagger}= \tilde{H}(D(g)\bm{k} +\bm{k}_0), \label{eq:UHU-node}
 \end{align} 
 where $\tilde{U}_{\bm{k}_0}(g) \equiv \tilde{V}_{D(g)\bm{k}_0 -\bm{k}_0}^{\dagger} \tilde{U}(g)$. Hence, nodes at $\bm{k}_0$ obey the symmetry operation $\tilde{U}_{\bm{k}_0}(g)$ rather than $\tilde{U}(g)$. Consider the commutation relation between $\{g_1|\bm{\tau}_1\}$ and $\{g_2|\bm{\tau}_2\}$ which belong to the little group of $\bm{k}_0$. The product of $\tilde{U}_{\bm{k}_0}(g_1)$ and $\tilde{U}_{\bm{k}_0}(g_2)$ is calculated as (see Appendix~\ref{app:Uk0})
  \begin{align}
  \tilde{U}_{\bm{k}_0}(g_1) \tilde{U}_{\bm{k}_0}(g_2) = \omega_{g_1,g_2}^{\bm{k}_0} \tilde{U}_{\bm{k}_0}(g_1g_2). \label{eq:Uk0-asso1}
 \end{align}
 Therefore, $\tilde{U}_{\bm{k}_0}$ satisfies the same relationship as Eq.~(\ref{eq:asso1}), implying that $V_{\bm{G}}$ gives the correct factor system. In addition, if $\tilde{U}_{\bm{k}_0}(g_1g_2) = t_{g_1,g_2}  \tilde{U}_{\bm{k}_0}(g_2g_1)$, we obtain
 \begin{align}
   \tilde{U}_{\bm{k}_0}(g_1) \tilde{U}_{\bm{k}_0}(g_2)=  t_{g_1,g_2} \alpha_{g_1,g_2}^{\bm{k}_0} \tilde{U}_{\bm{k}_0}(g_2) \tilde{U}_{\bm{k}_0}(g_1), \label{eq:Uk0-asso2}
 \end{align}
which coincides with Eq.~(\ref{eq:asso2}).

In closing this section, we remark a few properties of the factor system $\omega_{g_1,g_2}^{\bm{k}}$. In the case that $\{g_1 | \bm{\tau}_1 \}$ and $\{g_2 | \bm{\tau}_2 \}$ are an order-two operator, i.e., two-fold screw (rotation), glide (reflection), and spatial-inversion symmetries, $\alpha_{g_1,g_2}^{\bm{k}}$ of the factor system is simplified as
\begin{align}
 \alpha_{g_1,g_2}^{\bm{k}} = e^{i D(g_1) D(g_2) \bm{k} \cdot [(D(g_2) \bm{\tau}_1-\bm{\tau}_1)-(D(g_1) \bm{\tau}_2-\bm{\tau}_2)]}. \label{eq:F-system}
\end{align} 
In addition, PHS and TRS act trivially in real space, i.e., $D(C)=D(T)=\bm{1}_3$, so $\omega_{g,C}^{\bm{k}} = \omega_{g,T}^{\bm{k}}=1 $ for any $\{g|\bm{\tau}\} \in G$ and $\bm{k}$. In the following sections, Eqs.~(\ref{eq:Uk0-asso2}) and (\ref{eq:F-system}) are essential for nodes protected by nonsymmorphic symmetry.

\section{Line node in nonsymmorphc odd-parity superconductors}
\label{sec:line-node}

In this section, we revisit a line node in odd-parity SCs predicted by Micklitz and Norman~\cite{Norman:1995,Micklitz:2009}. As the minimal condition, consider a time-reversal invariant odd-parity SC with two-fold screw symmetry. The superconducting state possesses PHS $C$ ($C^2=1$), TRS $\mathcal{T}$ ($\mathcal{T}^2=-1$), IS $\mathcal{P}$ ($\mathcal{P}^2=1$ and $\{C,\mathcal{P}\}=0$), and the two-fold screw symmetry whose axis is perpendicular to a line node. When the screw axis is chosen to be the $z$-axis, the two-fold screw operator is algebraically described by $\left\{C_{2z}| \frac{1}{2} \hat{\bm{z}} \right\}$, where $\hat{\bm{z}}$ is a unit lattice vector along $z$-axis, and $C_{2z}$ is a two-fold rotation operator around the $z$-axis. The matrix representation of $\left\{C_{2z}| \frac{1}{2} \hat{\bm{z}} \right\}$ in Eq.(\ref{eq:GA}) is $D_{\bm{k}}\left (\left\{C_{2z}| \frac{1}{2} \hat{\bm{z}} \right\}\right) = e^{-i \frac{k_z}{2}} U(C_{2z})$. Combining the two-fold screw with $P$, a mirror-reflection operator is also defined as
\begin{align}
D_{\bm{k}}\left (\left\{PC_{2z} \Big| \frac{1}{2} \hat{\bm{z}} \right\}\right) &= D_{\bm{k}}\left (\left\{ M_{xy}\Big| \frac{1}{2} \hat{\bm{z}} \right\}\right) \notag \\
&= e^{-i \frac{k_z}{2}} U(M_{xy}), \label{eq:mirror-op}
\end{align} 
 where $M_{xy}$ is the mirror-reflection operator with respects to the $xy$ plane. Due to the spinor representation of rotation, we have $U(C_{2z})^2 = U(M_{xy})^2 = -1$. By calculating the factor system in Eq.~(\ref{eq:F-system}), the commutation relation between $D_{\bm{k}}\left (\left\{ M_{xy}| \frac{1}{2} \hat{\bm{z}} \right\}\right)$ and $P$ yields 
\begin{align}
 U(M_{xy}) P= e^{i k_z} P U(M_{xy}). \label{eq:P-M}
\end{align}
Here, we implicitly assume that $P$ commutes with $U(M_{xy})$ as usual. More general cases are discussed in Sec.~\ref{sec:general}. 
In what follows, we elucidate the existence of stable line nodes in odd-parity SCs in two different ways. In Sec.~\ref{subsec:Group},  we rely on the group theoretical method by focusing on the symmetry of Cooper pairs. Then, in Sec.~~\ref{subsec:Topology}, we develop a topological approach. 

\subsection{Group theoretical approach}
\label{subsec:Group}
\begin{figure}[tbp]
\centering
 \includegraphics[width=8cm]{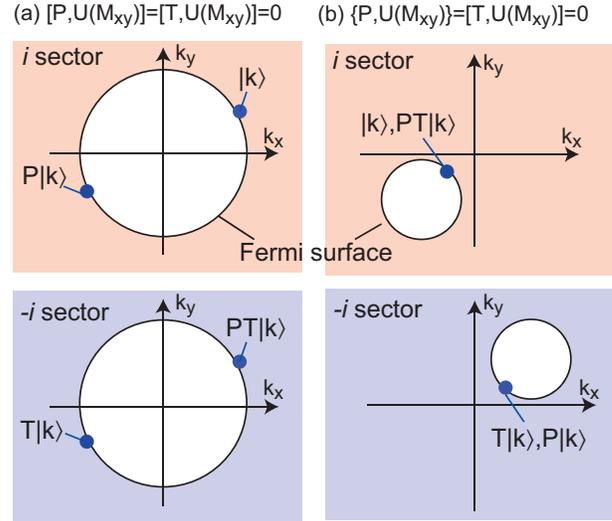}
 \caption{(Color online) Schematic illustration of the mirror-reflection eigenvalue of electron states, assuming that $U(M_{xy})|k\rangle = +i |k \rangle$. In the presence of TRS and IS, the Kramer's degeneracy occurs at any $\bm{k}$. The blue points indicate the electron states on the Fermi surface. $|k\rangle$ and $PT|k\rangle$ are the Kramer's pair located at $\bm{k}$ and $P|k\rangle$ and $T|k\rangle$ at $-\bm{k}$. If the mirror-reflection symmetry is present, each electronic state has the mirror-reflection eigenvalue $\pm i$ on the mirror-invariant plane, which is determined systematically according to the commutation relation between $P$, $T$, and $U(M_{xy})$. (a) and (b) describe the case of $[P,U(M_{xy})]=[T,U(M_{xy})]=0$ and of $\{P,U(M_{xy})\}=[T,U(M_{xy})]=0$, respectively. The upper (lower) figure represent the mirror-reflection eigenspace with $+i$ ($-i$). The Cooper pair has the mirror-reflection eigenvalue $+1$ (-1) if two electrons that form a Cooper pair lie in the different (same) eigenspace. Note that, in case (b), the Fermi surface needs a non-centrosymmetric shape in each mirror eigenspace in order to avoid a fully-gapped SC.}\label{fig:Cooper-pair}
\end{figure} 

In this subsection, we prove the existence of a stable line node at the BZ face based on the group theoretical method. The following argument essentially follows the Norman's one~\cite{Norman:1995} with simplification. When TRS and IS are present in a normal metal, the Kramer's doublet exists at an arbitrary $\bm{k}$, labeled by $|k\rangle$ and $PT |k \rangle$. Here, $|k\rangle$ represents an electronic state with momentum $\bm{k}$ and a pseudo spin $\beta$ (under $PT$: $\beta \to -\beta$). $|k\rangle$ and $PT |k \rangle$ correspond to spin-up and spin-down electronic states in the SOC free limit. Also, $P|k \rangle$ and $T|k \rangle$ describe the Kramer's doublet at $-\bm{k}$. 
When electrons at $\bm{k}$ and $-\bm{k}$ form a Cooper pair, we have a single even-parity pairing $(k,Tk) - (PTk,Pk)$ and three odd-parity pairings $(k,Pk)$, $(PTk,Tk)$, and $(k,Tk) + (PTk,Pk)$, where $(\cdot,\cdot)$ represents the electron pairs forming the Cooper pair. Introducing the $\bm{d}$-vector representation, each spin-triplet pairing is described by $-d_x+id_y$, $d_x + id_y$, and $d_z$, respectively. Here, we assume time-reversal invariant spin-triplet SCs that requires $\bm{d} \in \mathbb{R}^3$. In order to obtain a line node in the three-dimensional momentum space, it is necessary to fulfill $\bm{d} = \bm{0}$ along a curve on the Fermi surface. However, it is vanishingly improbable to satisfy the three conditions on the Fermi  surface at the same time. Thus, we need crystal symmetry. A line node may appear on a cross line between the Fermi surface and a higher symmetric plane where some of $\bm{d}$ vanishes, so we here consider mirror-reflection symmetry.  On the mirror-invariant plane, an electronic state $|k\rangle$ is an eigenstate of the mirror-reflection operator. Without loss of generality, we assume that $|k\rangle$ has the mirror-reflection eigenvalue $+i$. Then, the mirror-reflection eigenvalue of other electrons is systematically determined by the commutation relation between $T$, $P$, and $U(M_{xy})$. Likewise, the mirror-reflection eigenvalue of Cooper pairs is given by the product of that of two electrons and takes $\pm 1$. We have a mirror-reflection symmetry protected line node if mirror-reflection symmetry forces all components of $\bm{d}$-vector to vanish simultaneously on the mirror-invariant plane.

First, consider the mirror-reflection symmetry in Eq~(\ref{eq:mirror-op}) and the mirror-invariant plane at $k_z=0$. From Eq.~(\ref{eq:P-M}), $[P,U(M_{xy})] = 0$. We also have $[T,U(M_{xy})]=0$. With the anti-unitarity of $T$ in mind, $|k\rangle$ and $P|k\rangle$ take the mirror-reflection eigenvalue $+i$ and $T|k\rangle$ and $PT|k\rangle$ take $-i$. Thus, the Cooper pairs $(k,Tk)$ and $(PTk,Pk)$ have the mirror-reflection eigenvalue $+1$, whereas $(k,Pk)$ and $(PTk,Tk)$ have $-1$. (See Fig.~\ref{fig:Cooper-pair} (a)). Hence, when the Copper pair takes the mirror-reflection eigenvalue +1, only the $d_z$ component consisting of $(k,Tk)$ and $(PTk,Pk)$ survives on the mirror-invariant plane, while when the Cooper pair takes -1, the other $d_x$ and $d_y$ components are non-vanishing. That is, whichever mirror-reflection eigenvalue you take, $d_x$, $d_y$, and $d_z$ cannot vanish simultaneously, which means that mirror-reflection symmetry does not allow a line node in spin-triplet SCs. Accordingly, a line node at $k_z =0$ is unstable in time-reversal invariant spin-triplet SCs with and without mirror-reflection symmetry. This result is known as the Blount's theorem~\cite{Blount:1985}. 
 
 Next, consider the mirror-invariant plane at $k_z=\pi$. From Eq.~(\ref{eq:P-M}), we obtain $\{P,U(M_{xy})\}=0$ in addition to $[T,U(M_{xy})]=0$, leading to $+i$ for $|k\rangle$ and $PT|k\rangle$; $-i$ for $P|k\rangle$ and $T|k\rangle$. In contrast to the mirror-invariant plane at $k_z =0$, all of Cooper pairs $(k,Tk)$, $(PTk,Pk)$, $(k,Pk)$, and $(PTk,Tk)$ have the same mirror-reflection eigenvalue $+1$. (See Fig.~\ref{fig:Cooper-pair} (b)). 
That is, all components of $\bm{d}$-vector vanish simultaneously when the Cooper pair takes $-1$, leading to a stable line node at the BZ face~\cite{Norman:1995,Micklitz:2009}. The result does not contradict with the Blount's theorem since the commutation relation between $P$ and $U(M_{xy})$ changes at the BZ face.
To sum up, the mirror-reflection symmetry allows a symmetry protected line node only when $\{P,U(M_{xy})\}=[T,U(M_{xy})]=0$ and the Cooper pair is odd under the mirror-reflection operation. 

Whereas it is not clear in the original Norman's argument, the SOC is important to have a stable line node in odd-parity SCs. Without the SOC, there is four-fold degeneracy on the Fermi surface at $k_z =\pi$: As mentioned above, $|k\rangle$ and $PT|k \rangle$ have the same eigenvalue of $U(M_{xy})$ at $k_z=\pi$. Since $|k\rangle$ and $PT|k \rangle$ have the same momentum $\bm{k}$, there is two-fold degeneracy at each $\bm{k}$ in the $U(M_{xy})=i$ subsector. In the absence of SOC, on the other hand, spin is a good quantum number, so $|k \rangle$ and $PT | k\rangle$ can be written as $|k \uparrow \rangle$ and $PT |k \uparrow \rangle$, respectively. In this case, we also have full spin-rotation symmetry, which can flip the spin and the eigenvalue of $U(M_{xy})$ at the same time. Thus, using the spin-rotation symmetry, we obtain $|k \downarrow \rangle$ and $PT |k \downarrow \rangle$, which have the same energy and momentum as $|k \uparrow \rangle$ and $PT |k \uparrow \rangle$, but have the different eigenvalue $-i$ of $U(M_{xy})$. In total, we have four-fold degeneracy on the Fermi surface at $k_z=\pi$. 

Under this situation, we cannot have a stable line node in general. Because of the additional degeneracy, there are additional possible Cooper pairs $(k\uparrow, Tk\downarrow)$, $(PTk \uparrow,Pk \downarrow)$, $(k\uparrow, Pk\downarrow)$, and $(PTk\uparrow, Tk\downarrow)$, which take the mirror-reflection eigenvalue $-1$. Thus, even when the Cooper pair is odd under the mirror reflection, the $\bm{d}$-vector  of the additional Cooper pair survives at $k_z =\pi$. Consequently, no stable line node can be obtained. 
 
\subsection{Topological approach} 
\label{subsec:Topology}

Here we prove the stability of the line node from the topological point of view. We assume that line nodes exist at $k_z=0$ and $k_z=\pi$. Let $\tilde{H}(\bm{k})$ be the BdG Hamiltonian defined by Eq.~(\ref{eq:BdG2}). From Eq.~(\ref{eq:UHU}), the action of $\left\{M_{xy}| \frac{1}{2} \hat{\bm{z}} \right\}$ on the BdG Hamiltonian is
\begin{align}
 \tilde{U}(M_{xy}) \tilde{H} (k_x,k_y,k_z) \tilde{U} (M_{xy})^{\dagger} = \tilde{H} (k_x,k_y,-k_z). \label{eq:MHM}
\end{align}  
For mirror-reflection symmetry, Eq.~(\ref{eq:BdG-E}) becomes $U(M_{xy}) \Delta(k_x,k_y,k_z) U(M_{xy})^{\dagger} = \eta_{C,M} \Delta(k_x,k_y,-k_z) $, under which $\tilde{U}(M_{xy})=\diag [U(M_{xy}), \eta_{C,M} U(M_{xy})^{\ast}]$ obeys $C\tilde{U}(M_{xy})=\eta_{C,M}\tilde{U}(M_{xy})C$.
We label the position of a line node as $\bm{k}_M$ for $k_z=0$ and $\bm{k}_M'$ for $k_z=\pi$, which are invariant under the mirror-reflection operation up to an RL vector. From Eq.~(\ref{eq:UHU-node}), we have
\begin{align}
 &\tilde{U}(M_{xy}) \tilde{H} (\bm{k} +\bm{k}_M) \tilde{U} (M_{xy})^{\dagger} = \tilde{H} (D(M_{xy})\bm{k} + \bm{k}_M), \label{eq:BdG-0} \\
 &\tilde{U}_{\bm{k}_{M}'}(M_{xy}) \tilde{H} (\bm{k} +\bm{k}_{M}') \tilde{U}_{\bm{k}_{M}'} (M_{xy})^{\dagger}  = \tilde{H} (D(M_{xy})\bm{k} + \bm{k}_{M}'), \label{eq:BdG-pi}
\end{align}
where $\tilde{U}_{\bm{k}_M'}(M_{xy}) = V^{\dagger}_{-2\pi \hat{\bm{z}}}\tilde{U}(M_{xy}) = V_{2\pi \hat{\bm{z}}}\tilde{U}(M_{xy})$. It is found from Eqs.~(\ref{eq:Uk0-asso1}) and (\ref{eq:P-M}) that $V_{2\pi \hat{\bm{z}}}$ gives rise to the commutation relation, $[\mathcal{P},\tilde{U}(M_{xy})]=\{\mathcal{P},\tilde{U}_{\bm{k}_M'}(M_{xy})\} = 0$. Hence, the commutation relation between mirror reflection and inversion differs between the mirror-invariant planes at $\bm{k}_M$ and at $\bm{k}_M'$. As shown below, this difference is crucial for protecting a line node at the BZ face. 

To argue the topological stability of the line node, we employ the Clifford algebra extension method~\cite{Morimoto:2013,Shiozaki:2014,Morimoto:2014,Kobayashi:2014}, which leads to the correct topological classification of the gapped systems. For gapless nodes, one can consider a sphere or a circle enclosing the gapless nodes in the momentum space, instead of the whole BZ. Since the Hamiltonian of the nodes is fully gapped on the sphere or the circle, the topological classification of gapped systems is also applicable to the gapless nodes. Following Ref.~\onlinecite{Kobayashi:2014}, we expand the BdG Hamiltonian around a line node, yielding a massless Dirac Hamiltonian,
\begin{align}
 \tilde{H} (\bm{k} + \bm{k}_M) \simeq v_1k_z \gamma_1 + v_2 k_{\parallel} \gamma_2, \label{eq:H_LN}
\end{align}
where $k_\parallel$ is momentum parallel to the mirror-invariant plane and we neglect terms of order $k^n$ ($n >2$). 
The gamma matrices satisfy $\gamma_1^2 =\gamma_2^2=\bm{1}_m$ and $\{\gamma_1,\gamma_2\}=1$. The same expansion is applied to the BdG Hamiltonian at $\bm{k} + \bm{k}_M'$ as well. 
For the Hamiltonian expanded around the line node, a symmetry operation is relevant only when it does not change the position of the line node. PHS, TRS, and IS transfer $\bm{k}_M$ to $-\bm{k}_M$ ($\bm{k}'_M$ to $-\bm{k}'_M$) in the BZ, only their combinations are meaningful. For this reason, we consider the combined symmetry operators $C\mathcal{P}$ and $C\mathcal{T}$, where $\mathcal{P} \mathcal{T}$ is constructed from the combination of $C\mathcal{P}$ and $C\mathcal{T}$. Furthermore, the mirror-reflection operation~(\ref{eq:BdG-0}) or (\ref{eq:BdG-pi}) also does not change the position of the line node and is also relevant to the line node stability. 

For the massless Dirac Hamiltonian, gap-opening at nodes is equivalent to the existence of a mass term. Hence, the line node can be stable if Eq.~(\ref{eq:H_LN}) does not have any mass term under the symmetry constraint. The Clifford algebra extension method allows us to count the whole possible mass terms in Dirac Hamiltonians and clarify relevant topological numbers. Mathematically, the space of mass terms is described by the classifying space, $C_i$ ($i=0,1$) and $R_j$ ($j=0,\cdots 7$), and the topological number is defined by the zeroth homotopy group of the classifying space, $\pi_0(C_i) $ and $\pi_0 (R_j)$ (cf. Ref.~\onlinecite{Kobayashi:2014,Morimoto:2013} for more information). If the topological number is zero, a mass term create a gap, i.e., the line node is unstable, whereas if the topological number is nonzero, such a mass term is forbidden by topology and symmetry.  In preparation for the calculation, we define a set of Clifford algebras. We have the complex Clifford algebra $Cl_n = \{l_1,\cdots , l_n\}$ with $\{l_i,l_j\}=\delta_{ij}$ when the Hamiltonian does not have any anti-unitary symmetry, whereas we adopt the real Clifford algebra $Cl_{p,q} = \{e_1,\cdots,e_p;e_{p+1},\cdots,e_{p+q}\}$ when the Hamiltonian has anti-unitary symmetry, where $e_i$ is a Clifford algebra satisfying $\{e_i,e_j\} = 0 $ $(i\neq j)$, and $e_i^2=-1$ ($1\le i \le p$) and $e_i^2=+1$ ($p+1 \le i \le q$). $l_i$ and $e_i$ are constructed from $\gamma_i$'s and symmetry operators for the underlying Hamiltonian. In addition, in order to input the imaginary number $i$ in the real Clifford algebra, we introduce a generator $J$ ($J^2=-1$), which anti-commutes only with anti-unitary operators. 

 For illustration purpose, we first examine the line node stability in odd-parity SCs without assuming mirror-reflection symmetry. Odd-parity superconductivity implies the anti-commutation relation $\{C,\mathcal{P}\}=0$ with $\mathcal{P}^2=1$. In the case of TRS breaking odd-parity SCs, we have $\gamma_1$, $\gamma_2$, $J$, and $CP$. Adjusting the anti-commutation relation between them, the set of Clifford algebra is constructed as $Cl_{2,2}=\{C\mathcal{P},JC\mathcal{P};\gamma_1,\gamma_2\}$, where $(C\mathcal{P})^2 =(JC\mathcal{P})^2=-1$. According to Ref.~\onlinecite{Kobayashi:2014}, we calculate the  Clifford algebra extension problem in terms of $\gamma_2$, leading to $Cl_{2,1} \to Cl_{2,2}$ and the classifying space $R_{7}$. Since $\pi(R_7) =0$~\cite{Morimoto:2013}, a line node is topologically unstable. On the other hand, in the case of time-reversal invariant odd-parity SCs, we need to add $C\mathcal{T}$ in the above set. Hence, the set of Clifford algebra becomes $Cl_{3,2}=\{C\mathcal{P},JC\mathcal{P},C\mathcal{T};\gamma_1,\gamma_2\}$, where $(C\mathcal{T})^2 =-1$. The  Clifford algebra extension is $Cl_{3,1} \to Cl_{3,2}$, resulting in $R_{6}$ and $\pi_0(R_6)= 0$. Thus, a line node is topologically unstable as well. These results imply that an additional symmetry is necessary to stabilize a line node in odd-parity SCs with and without TRS. In what follows, we calculate the Clifford algebra extension problem in time-reversal invariant odd-parity SCs with mirror-reflection symmetry and compare it with the group theoretical results. 

 First, consider a line node in the mirror-invariant plane at $k_z =0$, in which the mirror-reflection operator satisfies $[\mathcal{P},\tilde{U}(M_{xy})] = [\mathcal{T},\tilde{U}(M_{xy})]=0$. The commutation relation between $U(M_{xy})$ and the combined operators becomes 
 \begin{subequations}
 \begin{align}
 &C\mathcal{P} \tilde{U}(M_{xy}) = \eta_{C,M} \tilde{U}(M_{xy}) C \mathcal{P}, \label{eq:CPM-0} \\
 &C\mathcal{T} \tilde{U}(M_{xy}) = \eta_{C,M} \tilde{U}(M_{xy}) C\mathcal{T}. \label{eq:CTM-0}
\end{align}
\end{subequations}
 
In the presence of mirror-reflection symmetry, we have $\gamma_1$, $\gamma_2$, $J$, $C\mathcal{P}$, $C\mathcal{T}$, and $\tilde{U}(M_{xy})$ as candidates of the Clifford algebra. Taking into account the sign of $\eta_{C,M}$, these algebras are packed in the set of Clifford algebras as $Cl_{3,3}=\{C\mathcal{P},JC\mathcal{P},C\mathcal{T};\gamma_1,\gamma_2,\gamma_1 \tilde{U}(M_{xy})\}$ for $\eta_{C,M}=+1$ and $Cl_{3,2} \otimes Cl_{0,1}=\{C\mathcal{P},JC\mathcal{P},C\mathcal{T};\gamma_1,\gamma_2\} \otimes \{ ;JC\mathcal{T}\tilde{U}(M_{xy}) \}$ for $\eta_{C,M}=-1$, where $JC\mathcal{T}\tilde{U}(M_{xy})$ with  $[JC\mathcal{T}\tilde{U}(M_{xy})]^2=+1$ commutes with the other Clifford algebras and does not affect the extension problem. Calculating the  Clifford algebra extension problem in terms of $\gamma_2$, we obtain $\pi_0 (R_7) = 0$ for $\eta_{M,C}=+1$ and  $\pi_0 (R_6) = 0$ for $\eta_{C,M}=-1$. Therefore, a line node is topologically unstable in both cases. As a result, mirror-reflection symmetry cannot stabilize a line node in time-reversal invariant odd-parity SCs. This result, together with the result without mirror-reflection symmetry, is the topological version of the Blount's theorem~\cite{Kobayashi:2014}.   

Next, consider a line node at $k_z = \pi$. Taking into account the effect of $V_{2\pi \hat{\bm{z}}}$, we replace $\tilde{U}(M_{xy})$ with $\tilde{U}_{\bm{k}'_M}(M_{xy})$. Then, the nontrivial factor system changes the commutation relation between IS and mirror-reflection symmetry operator, so we obtain
\begin{subequations}
\begin{align}
 &C\mathcal{P}\tilde{U}_{\bm{k}'_M}(M_{xy}) = -\eta_{C,M}\tilde{U}_{\bm{k}'_M}(M_{xy}) C\mathcal{P}, \label{eq:CPM-pi}  \\
 &C\mathcal{T} \tilde{U}_{\bm{k}'_M}(M_{xy}) = \eta_{C,M} \tilde{U}_{\bm{k}'_M}(M_{xy}) C\mathcal{T}. \label{eq:CTM-pi} 
\end{align}
\end{subequations}
Hence, the commutation relation with $C\mathcal{P}$ changes.  For $\eta_{C,M} = +1$, the set of Clifford algebras is given by $Cl_{4,2} = \{C\mathcal{P},JC\mathcal{P},C\mathcal{T},J\gamma_1 \tilde{U}_{\bm{k}'_M}(M_{xy}) ;\gamma_1,\gamma_2 \}$ and the Clifford algebra extension becomes $Cl_{4,1} \to Cl_{4,2}$, leading to $R_5$ and $\pi_0 (R_5)=0$. Thus, a line node is topologically unstable. On the other hand, for $\eta_{C,M} = -1$, the set of Clifford algebras is constructed as $Cl_{3,2} \otimes Cl_{1,0}=\{C\mathcal{P},JC\mathcal{P},C\mathcal{T};\gamma_1,\gamma_2\} \otimes \{C\mathcal{T}\tilde{U}_{\bm{k}'_M}(M_{xy}) ;\}$. Here, $C\mathcal{T}\tilde{U}_{\bm{k}'_M}(M_{xy})$ with $[C\mathcal{T}\tilde{U}_{\bm{k}'_M}(M_{xy})]^2 =-1$ commutes with the other Clifford algebras and thus plays a role of the complex factor. Then, the set of Clifford algebras changes to the complex case~\cite{Morimoto:2013}, $Cl_{3,2} \otimes Cl_{1,0} \simeq Cl_{5}$.  The Clifford algebra extension becomes $Cl_4 \to Cl_5$, leading to $C_0$ and $\pi_0 (C_0) = \mathbb{Z}$. As a result, a line node at the BZ face can be topologically stable for mirror-odd (and odd-parity) pairing states. 
This result reproduces the Norman's one, in spite that the argument is completely different. In Sec.~\ref{subsec:UPt3}, based on a recently proposed model of UPt$_3$, we show that the line node at the BZ face actually has a non-trivial topological number.

In the topological approach, we can generalize the above result to TRS breaking odd-parity SCs. The Clifford algebra extension is given by removing $C \mathcal{T}$ from the set of Clifford algebras. On the mirror-invariant plane at $k_z=0$, the set of Clifford algebras is constructed as $Cl_{2,3}=\{C\mathcal{P},JC\mathcal{P};\gamma_1,\gamma_2,\gamma_1 \tilde{U}(M_{xy})\}$ for $\eta_{C,M}=+1$ and $Cl_{3,2}=\{C\mathcal{P},JC\mathcal{P},J\gamma_1\tilde{U}(M_{xy});\gamma_1,\gamma_2\}$ for $\eta_{C,M}=-1$.  From the Clifford algebra extension problem in terms of $\gamma_2$, we obtain $\pi_0 (R_0) = \mathbb{Z}$ for $\eta_{C,M}=+1$ and  $\pi_0 (R_6) = 0$ for $\eta_{C,M}=-1$. Thus, a line node at $k_z=0$ is topologically stable when $\eta_{C,M}=+1$. On the other hand, on the mirror-invariant plane at $k_z=\pi$, the mirror-reflection operator is $\tilde{U}_{\bm{k}'_M}(M_{xy})$, which obeys $C\mathcal{P}\tilde{U}_{\bm{k}'_M}(M_{xy}) = -\eta_{C,M} \tilde{U}_{\bm{k}'_M}(M_{xy})C\mathcal{P}$. Since only difference between the above commutation relation and Eq.~(\ref{eq:CPM-0}) is minus sign in $\eta_{C,M}$, we can obtain the topological structure at $k_z=\pi$ from that at $k_z=0$ by changing the sign of $\eta_{C,M}$. Therefore, a line node at $k_z =\pi$ can be topologically stable when $\eta_{C,M}=-1$. In conclusion, there exists a topologically stable line node at $k_z=\pi$ regardless of TRS when the Cooper pair is odd under the mirror-reflection operation. It should be noted here that the possibility of stable line nodes at $k_z=0$ in the above is overlooked in the group theoretical method: Although only the original mirror-reflection symmetry exists at $k_z=0$, there may exist a stable line node. An example of the stable line node at $k_z=0$ has been given in Appendix 3 in Ref.~\onlinecite{Kobayashi:2014}. This result suggests that the topological approach is more powerful than the group theoretical method.

Finally, we present the topological number of nodal rings in the mirror-invariant plane. (Generally, a line node on a plane forms a nodal ring.) As shown in the above, a nodal ring is characterized by an integer. The topological number on the mirror-invariant plane is defined by
\begin{align}
 \mathcal{Q}_{\lambda} \equiv n_{\textrm{occ},\lambda}^{>}- n_{ \textrm{occ},\lambda}^{<} \in \mathbb{Z}, \label{eq:top-Q}
\end{align} 
where $n_{ \textrm{occ},\lambda}^{>}$ ($n_{ \textrm{occ},\lambda}^{<}$) is the number of the occupied states with mirror-reflection eigenvalue $\lambda$ outside (inside) a nodal ring. We readily verify that $\mathcal{Q}_{\lambda}$ is nontrivial only if $[C\mathcal{P},\tilde{U}(M_{xy})] =0$ in TRS breaking odd-parity SCs or $[C\mathcal{P},\tilde{U}(M_{xy})] =\{ C\mathcal{T},\tilde{U}(M_{xy}) \} = 0$ in time-reversal invariant odd-parity SCs. This is because $C\mathcal{P}$ symmetry leads to $n_{  \textrm{occ}, \lambda}^{> (<)} = N_{\lambda} - n_{  \textrm{occ}, \lambda}^{> (<)} $ when $\{C\mathcal{P}, \tilde{U}(M_{xy})\}=0$. ($N_{\lambda}$ is the total number of eigenstates with $\lambda$ and dose not depend on $\bm{k}$.) This means that $n_{  \textrm{occ}, \lambda}^{>} = n_{  \textrm{occ}, \lambda}^{<} = \frac{N_{\lambda}}{2}$, leading to $\mathcal{Q}_{\lambda}=0$. In the same way, $C\mathcal{T}$ symmetry leads to $\mathcal{Q}_{\lambda}=0$ when $[ C\mathcal{T},\tilde{U}(M_{xy}) ]=0$. Applying $\mathcal{Q}_{\lambda}$ to time-reversal invariant odd-parity SCs, the commutation relations at $k_z=0$ are given by Eqs.~(\ref{eq:CPM-0}) and (\ref{eq:CTM-0}), so $\mathcal{Q}_{\lambda}$ is always trivial. On the other hand, the commutation relations at $k_z =\pi$ are given by Eqs.~(\ref{eq:CPM-pi}) and (\ref{eq:CTM-pi}). That is, $\mathcal{Q}_{\lambda}$ is nontrivial only when $\eta_{C,M}=-1$. Accordingly, $\mathcal{Q}_{\lambda}$ coincides with the above argument. Note that the absence of SOC leads to $\mathcal{Q}_{\lambda} =0$ even when $\eta_{C,M}=-1$. (See Appendix~\ref{app:vanish}.)

 \subsection{Application to superconducting state in UPt$_3$}
 \label{subsec:UPt3}
 \begin{figure}[tbp]
\centering
 \includegraphics[width=8.5cm]{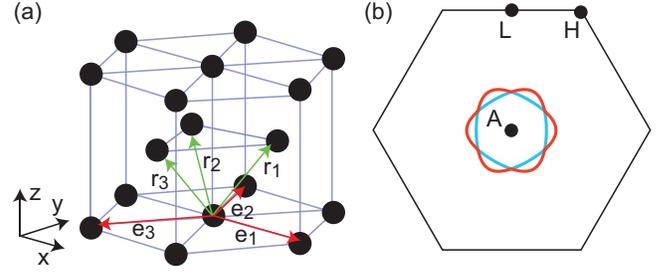}
 \caption{(Color online) (a) Crystal structure of UPt$_3$, where black points indicate the position of U ions~\cite{Yanase:2016}. (b) Fermi surfaces at $k_z=\pi$ from the tight-binding Hamiltonian (\ref{eq:UPt3-H}) with the parameters $(t,t_z,t',\alpha,\mu) = (1,-4,1,2,12)$. The red and blue lines represent doubly-degenerate Fermi surfaces around the A point. If $\alpha=0$, these Fermi surfaces are overlapped and cause fourfold degeneracy. }\label{fig:UPt3}
\end{figure} 

 We demonstrate the nonsymmorphic symmetry protected line node concretely in the tight-binding model for $E_{2u}$-superconducting state of UPt$_3$ B-phase~\cite{Yanase:2016}. The BdG Hamiltonian is given by
 \begin{subequations} 
 \begin{align}
  \mathcal{E}_{mm'ss'}(\bm{k}) &= \xi (\bm{k}) \delta_{m,m'}\delta_{s,s'}+a_{mm'}(\bm{k}) \delta_{s,s'} \notag \\ 
  &+ (-1)^{3-m} \alpha \bm{g}(\bm{k}) \cdot \bm{s}_{ss'} \delta_{m,m'}, \label{eq:UPt3-H} \\ 
  \Delta_{mm'ss'}(\bm{k}) & = \frac{\Delta}{\sqrt{2}} (\Gamma^a_{mm'ss'} (\bm{k}) + i \Gamma^b_{mm'ss'} (\bm{k})), \label{eq:UPt3-delta}
 \end{align}
 \end{subequations}
 where $m=1,2$ and $s=\uparrow,\downarrow$ are indexes of sublattice and spin, respectively. $\Gamma^a(\bm{k})$ and $\Gamma^b (\bm{k})$ represent the order parameter in the superconducting state of a two-dimensional irreducible representation $E_{2u}$. Taking into account the local violation of inversion symmetry, which gives rise to the sublattice-dependent Zeeman type SOC~\cite{Fischer:2011,Maruyama:2012}, each term in the normal Hamiltonian is described as
 \begin{subequations} 
 \begin{align}
  &\xi (\bm{k}) = 2 t \sum_{i=1}^3 \cos \bm{k}_{\parallel} \cdot \bm{e}_i + 2 t_z \cos k_z - \mu, \\
  & a_{11}=a_{22}=0, \\
  &a_{12}(\bm{k}) = a_{21}(\bm{k})^{\ast} = 2 t' \cos \frac{k_z}{2} \sum_{i=1}^3 e^{i \bm{k}_{\parallel} \cdot \bm{r}_i}, \\
  &\bm{g}(\bm{k}) = \hat{\bm{z}} \sum_{i=1}^3 \sin \bm{k}_{\parallel} \cdot \bm{e}_i,
 \end{align}
 \end{subequations}
with $\bm{k}_{\parallel} = (k_x,k_y,0)$. As in Fig.~\ref{fig:UPt3} (a), $\bm{e}_1 = (1,0,0)$, $\bm{e}_2 = \left( -\frac{1}{2}, \frac{\sqrt{3}}{2},0\right)$, and $\bm{e}_3 = \left( -\frac{1}{2}, -\frac{\sqrt{3}}{2},0\right)$ are unit vectors in the two-dimensional triangular lattice and $\bm{r}_1 = \left( \frac{1}{2}, \frac{1}{2\sqrt{3}},\frac{1}{2}\right) $, $\bm{r}_2 = \left( - \frac{1}{2}, \frac{1}{2\sqrt{3}},\frac{1}{2}\right) $, and $\bm{r}_3 = \left( 0, -\frac{1}{\sqrt{3}},\frac{1}{2}\right) $ are non-primitive lattice vectors in two dimension. The symmetry allowed gap function is constructed from
 \begin{subequations} 
 \begin{align}
  \Gamma^a (\bm{k}) &= [\delta \{p_x (\bm{k}) s_x - p_y (\bm{k}) s_y \} \sigma_0 \notag \\
                 & +f_{(x^2-y^2)z} (\bm{k}) s_z \sigma_x -d_{yz} (\bm{k}) s_z \sigma_y] i s_y \\
   \Gamma^b (\bm{k}) &= [\delta \{p_x (\bm{k}) s_x + p_y (\bm{k}) s_y \} \sigma_0 \notag \\
                 & +f_{xyz} (\bm{k}) s_z \sigma_x -d_{xz} (\bm{k}) s_z \sigma_y] i s_y,               
 \end{align}
 \end{subequations}  
 where $s_{\alpha}=(\bm{1}_2,\bm{s})$ and $\sigma_{\alpha}=(\bm{1}_2,\bm{\sigma})$ are the identity and Pauli matrices in the spin and sublattice spaces. The $p$-wave, $f$-wave, and $d$-wave components of the basis function are
 \begin{subequations}
 \begin{align}
  &p_x(\bm{k}) = \sum_{i=1}^3 e_i^x \sin \bm{k}_{\parallel} \cdot \bm{e}_i, \\
  &p_y(\bm{k}) = \sum_{i=1}^3 e_i^y \sin \bm{k}_{\parallel} \cdot \bm{e}_i, \\
  &f_{(x^2-y^2)z} (\bm{k}) = - \sin \frac{k_z}{2} {\rm Re} \left[ \frac{1}{2} e^{i\bm{k}_{\parallel} \cdot \bm{r}_1}+\frac{1}{2} e^{i\bm{k}_{\parallel} \cdot \bm{r}_2} -e^{i\bm{k}_{\parallel} \cdot \bm{r}_3} \right], \\
  &f_{xyz}(\bm{k}) = -\frac{\sqrt{3}}{2} \sin \frac{k_z}{2} {\rm Re} [e^{i\bm{k}_{\parallel} \cdot \bm{r}_1} -e^{i\bm{k}_{\parallel} \cdot \bm{r}_2}], \\
  &d_{yz} (\bm{k}) = - \sin \frac{k_z}{2} {\rm Im} \left[  \frac{1}{2} e^{i\bm{k}_{\parallel} \cdot \bm{r}_1}+\frac{1}{2} e^{i\bm{k}_{\parallel} \cdot \bm{r}_2} -e^{i\bm{k}_{\parallel} \cdot \bm{r}_3} \right], \\
  &d_{xz} (\bm{k}) = -\frac{\sqrt{3}}{2} \sin \frac{k_z}{2} {\rm Im}  [e^{i\bm{k}_{\parallel} \cdot \bm{r}_1} -e^{i\bm{k}_{\parallel} \cdot \bm{r}_2}].
 \end{align}   
 \end{subequations}
As shown in Ref.~\onlinecite{Yanase:2016}, there exist six stable nodal rings at the BZ face when parameters are set in such a way that the Fermi surfaces appear around the A point, and the effect of the Zeeman-type SOC is included. (See Fig.~\ref{fig:UPt3} (b)). We point out in the following that the obtained nodal rings have nontrivial topological numbers owing to nonsymmorphic symmetry. 
 
\begin{figure}[tbp]
\centering
 \includegraphics[width=7cm]{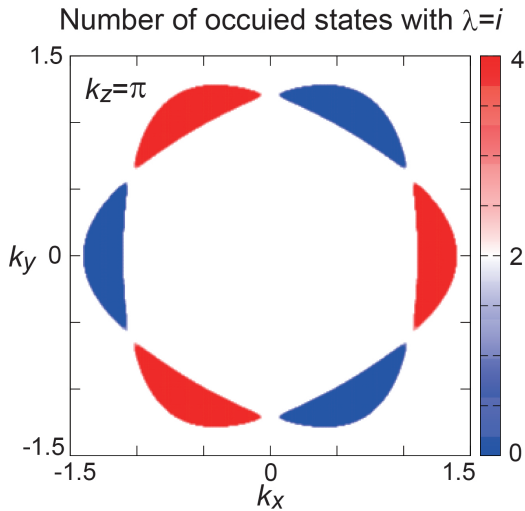}
 \caption{(Color online) Number of occupied states with mirror-reflection eigenvalue $+i$ as a function of $(k_x,k_y)$, which is calculated by numerically diagonalizing the BdG Hamiltonian (\ref{eq:UPt3-H}) and (\ref{eq:UPt3-delta}) with the parameters $(t,t_z,t',\alpha,\mu,\Delta,\delta) = (1,-4,1,2,12,0.1,0.04)$. The red and blue regions indicate the inside of nodal rings, indicating that they are protected by the mirror topological number $\mathcal{Q}_{\lambda} = \pm 2$.}\label{fig:UPt3-LN}
\end{figure}  
 
 The symmetries relevant to the stability of the nodal rings are $C=\sigma_0 s_0 \tau_x K$, $\mathcal{P}= \sigma_x s_0 \tau_z$, and $\tilde{U}(M_{xy}) = i \sigma_0 s_z \tau_0$, where $\tau_{\alpha} = (\bm{1}_2 , \bm{\tau})$ describes the identity and Pauli matrices in the Nambu space. At this point, the symmetry operators satisfy $\{ CP, \tilde{U} (M_{xy})\}=0$, namely, there is no symmetry protected line node since $\mathcal{Q}_{\lambda} = 0$.  
 Following UPt$_3$ having nonsymmorphic space group $P6_3/mmc$, the BdG Hamiltonian satisfies Eq.~(\ref{eq:plusG}) with the non-primitive lattice vector $\bm{\tau} = \left( 0, -\frac{1}{\sqrt{3}}, \frac{1}{2}\right)$. It follows that 
 \begin{align}
  \mathcal{E} (\bm{k} +\bm{G}) = V_{\bm{G}} \mathcal{E}(\bm{k}) V_{\bm{G}} ^{\dagger}, \\
   \Delta (\bm{k} +\bm{G}) = V_{\bm{G}} \Delta(\bm{k}) V_{\bm{G}}^{\dagger} ,
 \end{align}
 with $\bm{G}=m_1 \bm{b}_1 +m_2 \bm{b}_2 +m_3 \bm{b}_3$ ($m_1,m_2,m_3 \in \mathbb{Z}$) and 
 \begin{align}
  V_{\bm{G}} = \begin{pmatrix} 1 &0 \\ 0 & e^{-i \bm{G} \cdot \bm{\tau}} \end{pmatrix}_{\sigma} \otimes s_0 . \label{eq:UPt3-V}
 \end{align}
 Here, $\bm{b}_1 = \frac{4 \pi}{\sqrt{3}}\left( \frac{\sqrt{3}}{2},\frac{1}{2} ,0 \right)$, $\bm{b}_2 = \frac{4 \pi}{\sqrt{3}} (0,1 ,0 )$, and $\bm{b}_3 = 2\pi(0,0 ,1)$ are the RL vectors.
 Therefore, using Eq.~(\ref{eq:BdG-pi}), the mirror-reflection operator at $k_z=\pi$ is 
 \begin{align}
 \tilde{U}_{\bm{k}_M'} (M_{xy})  = \tilde{V}_{2 \pi \hat{\bm{z}}} \tilde{U} (M_{xy}) = i \sigma_z s_z \tau_0. \label{eq:NSmirror}
 \end{align}
 Thus, the mirror-reflection operator satisfies $[C\mathcal{P},\tilde{U}_{\bm{k}_M'} (M_{xy})]=0$ at $k_z=\pi$, which is attributed to the nontrivial factor system between $P$ and $\left\{ U(M_{xy})|\frac{1}{2}\hat{\bm{z}} \right\}$. To verify that the nodal rings are protected by the topological number, we evaluate $\mathcal{Q}_{+i}$ in the mirror-invariant plane at $k_z=\pi$. In Fig.~\ref{fig:UPt3-LN}, we show the number of occupied states with the mirror-reflection eigenvalue $+i$ by numerically diagonalizing the BdG Hamiltonian. The red and blue regions indicate the inside of nodal rings. By calculating $Q_{+i}$ according to the definition (\ref{eq:top-Q}), we find that the red and blue nodal rings have $\mathcal{Q}_{+i}=-2$ and $+2$, respectively. That is, the six nodal rings are topologically protected with help from nonsymmorphic symmetry. It must be noted that the nodal rings disappear when $\alpha=0$, implying that the Zeeman-type SOC plays an important role to protect the nodal rings. 
 
More generally, the heavy fermion superconductor UPt$_3$ exhibits multiple superconducting phases in the field-temperature phase~\cite{Joynt:2002}. The order parameter that covers the enter range of superconducting phases is described by $\Delta(\bm{k}) = \Delta[\eta_a \Gamma^a(\bm{k}) + \eta_b \Gamma^b (\bm{k})]$ with $(\eta_a,\eta_b) = (1,i\eta)/\sqrt{1+\eta^2}$ ($\eta \in \mathbb{R}$). By tuning a real parameter $\eta$, we obtain the A phase ($\eta= \infty$), the B phase ($0< \eta < \infty$), and the C phase ($ \eta = 0$). In contrast to the B phase, the A and C phases recover TRS. Nevertheless, as discussed in Sec~\ref{subsec:Topology}, a nonsymmorphic symmetry protected line node is stable independent of TRS. Thus, the nodal rings are robust for any $\eta$, as long as the mirror reflection symmetry (\ref{eq:NSmirror}) and the Zeeman-type SOC exists.  
 
 \section{Generalization to other systems}
 \label{sec:general}
 
 \begin{table*}[tbp]
\centering
\caption{Classification of nodes under PHS, TRS, IS, and mirror-reflection symmetry~\cite{Kobayashi:2014}. The first and second columns represent the symmetry operations and the parity of the gap function. The third column lists the commutation relation between $C\mathcal{P}$, $C\mathcal{T}$, and $\tilde{M}$, where $\tilde{M}^{\eta_{CP,M}}$ ($\tilde{M}^{\eta_{CP,M},\eta_{CT,M}}$) represents time-reversal breaking (invariant) case. 
The forth column shows the corresponding classifying spaces. The following columns represent the topological numbers for each codimension. \vspace*{2mm}} \label{tab:reflection-node}

\begin{tabular}{ccccccc}
\hline \hline
 Symmetry  & Parity & Mirror            & Classifying space  & $p=0$ & $p=1$ & $p=2$  \\ \hline
 \{$C\mathcal{P},\tilde{M}$\} & Even & $\tilde{M}^+$ &  $ R_{p+3} $ & $0$ & $2\mathbb{Z}$ & $0$ \\
                  &           & $\tilde{M}^-$   & $ R_{p+1}$ & $\mathbb{Z}_2$ & $\mathbb{Z}_2$ & $0$ \\
                  &  Odd  & $\tilde{M}^{+}$&  $ R_{p-1} $ & $0$ & $\mathbb{Z}$ & $\mathbb{Z}_2$ \\
                 &         &$\tilde{M}^-$ & $ R_{p-3}$ & $0$ & $0$ & $0$ \\ \hline               
\{$C\mathcal{P},C\mathcal{T},\tilde{M}$\} & Even &$\tilde{M}^{++}$ & $R_{p+4}$& $2\mathbb{Z}$ & $0$ & $0$ \\
                      &          & $\tilde{M}^{-+}$& $R_{p+2}$& $\mathbb{Z}_2$ & $0$ & $2\mathbb{Z}$ \\              
              &       &        $\tilde{M}^{+-}$& $R_{p+3}$ & $0$ & $2\mathbb{Z}$& $0$ \\
                       &         & $\tilde{M}^{--}$& $C_{p+1}$ & $0$ & $\mathbb{Z}$ & $0$ \\
           & Odd   & $\tilde{M}^{++}$&  $ R_{p-2} $ & $0$ & $0$ & $\mathbb{Z}$ \\
          &     &$\tilde{M}^{-+}$& $ R_{p-4} $ & $2\mathbb{Z}$ & $0$ & $0$ \\
         &      &$\tilde{M}^{+-}$& $ C_{p+3}$ & $0$ & $\mathbb{Z}$ & $0$  \\
           &         & $\tilde{M}^{--}$& $ R_{p-3}$ & $0$ & $0$ & $0$  \\ \hline \hline
\end{tabular}

 \end{table*}
 
 So far, we discussed nonsymmorphic symmetry protected line nodes bearing the Norman's discussion in mind. An advantage of the topological approach is that one can generalize the argument to other nodes and other symmetry classes systematically.  
 
 For this purpose, we consider a generic node at $\bm{k}_0$ described by the massless Dirac Hamiltonian,
 \begin{align}
 \tilde{H} (\bm{k} + \bm{k}_0) \simeq \sum_{i=1}^{p+1} v_i k_i \gamma_i, 
 \end{align} 
 where $k_i$'s are momentum on a $p$-dimensional sphere enclosing the node, the gamma matrices satisfy $\{\gamma_i, \gamma_j \} = 0$ ($i \neq j$), and $p$ specifies the transverse dimension of nodes, which we call codimension of nodes (Codim. for short). For example, $p=0$, $p=1$, and $p=2$ represent a gapless superconductor (a surface node), a line node, and a point node in the three-dimensional momentum space.  
 
 For symmetries protecting the node, we consider PHS, TRS, IS, and the mirror reflection symmetry $M$. (Without loss of generality, we assume $M^2=-1$.) In a manner similar to Sec.~\ref{subsec:Topology}, topological stability of the node depends on the commutation relation between $C\mathcal{P}$, $C\mathcal{T}$, and $\tilde{M}$, where $\tilde{M}=\diag[M,\eta_{C,M} M^{\ast}]$ is the mirror-reflection operator in the Nambu space. Introducing $\eta_{S,M}$ to specify the commutation relation between $S (=C,\mathcal{P},\mathcal{T})$ and $\tilde{M}$ as $S\tilde{M} = \eta_{S,M}\tilde{M}S$, the commutation relation between $C\mathcal{P}$, $C\mathcal{T}$, and $\tilde{M}$ are given by $\eta_{CP,M}=\eta_{C,M} \eta_{P,M}$ and $\eta_{CT,M}=\eta_{C,M} \eta_{T,M}$. We label $\tilde{M}$ with these commutation relations as $\tilde{M}^{\eta_{CP,M}}$ ($\tilde{M}^{\eta_{CP,M},\eta_{CT,M}}$) for time-reversal breaking (invariant) case. In addition, the parity of the gap function takes either even-parity ($[C,\mathcal{P}]=0$) or odd-parity ($\{C,\mathcal{P}\}=0$) for each mirror symmetry class. Solving the the Clifford algebra extension problem in terms of $\gamma_{p+1}$ systematically~\cite{Kobayashi:2014}, we obtain the corresponding classifying space and the topological number for each Codim., as shown in Table~\ref{tab:reflection-node}.
 
 Finally, we take into account the influence of the factor system on the topological classification. To this end, we define a nonsymmorphic mirror operator as the combination of spatial inversion $P$ and two-fold screw symmetries $\{C_{2x_{\perp}}| \bm{\tau_{\perp}} \}$, $\{PC_{2x_{\perp}}| \bm{\tau}_{\perp} \} \equiv \{M| \bm{\tau}_{\perp} \}$, where $C_{2x_{\perp}}$ is a two-fold rotation operator in terms of the $x_{\perp}$ axis and $\bm{\tau}_{\perp}$ is a non-primitive lattice vector along the $x_{\perp}$ axis ($2 \bm{\tau}_{\perp}$ is a primitive lattice vector). As discussed in the previous section, only the nontrivial factor system changes the commutation relation between $\tilde{M}$ and $C\mathcal{P}$, which causes the change of mirror-reflection symmetry between the mirror-invariant plane at $k_{\perp} =0$ and at $k_{\perp} =\pi$:
 \begin{align}
  \begin{array}{@{\,} c|ccc @{\,}}
  {\rm Symmetry} & k_{\perp} =0    &      & k_{\perp} = \pi  \\ \hline
   \{C\mathcal{P},\tilde{M}\} &\tilde{M}^{\eta_{CP,M}} &\Longrightarrow &\tilde{M}^{-\eta_{CP,M}}  \\
   \{C\mathcal{P},C\mathcal{T},\tilde{M}\} & \tilde{M}^{\eta_{CP,M}, \eta_{CT,M}} &\Longrightarrow  &\tilde{M}^{-\eta_{CP,M}, \eta_{CT,M}} 
  \end{array} \label{eq:main}
 \end{align}
 Comparing Table~\ref{tab:reflection-node} with Eq.~(\ref{eq:main}), a nonsymmorphic symmetry protected line node in odd-parity SCs with and without TRS is classified by
 \begin{align}
  \begin{array}{@{\,} c|c|ccc @{\,}}
  {\rm Parity}& {\rm Codim.}&k_{\perp} =0    &      & k_{\perp} = \pi  \\ \hline
  {\rm Odd} & p=1 &\tilde{M}^{-} &\Longrightarrow  &\tilde{M}^{+} \\ 
  {\rm Odd} & p=1 &\tilde{M}^{--} &\Longrightarrow  &\tilde{M}^{+-} 
  \end{array} \label{eq:NS-line}
 \end{align}
 A line node in both cases is protected by the $\mathbb{Z}$ topological number on the BZ face and is characterized by $\mathcal{Q}_{\lambda}$. The $E_{2u}$ superconducting state of UPt$_3$ B-phase belongs to the first line in Eq.~(\ref{eq:NS-line}). Furthermore, provided that $[M,T]=[M,P]=0$ with spin-singlet or spin-triplet SCs in mind, we find two types of nonsymmorphic symmetry protected point nodes in Table~\ref{tab:reflection-node} as follows.
  \begin{align}
  \begin{array}{@{\,} c|c|ccc @{\,}}
  {\rm Parity}& {\rm Codim.}&k_{\perp} =0    &      & k_{\perp} = \pi  \\ \hline
  {\rm Odd} & p=2 &\tilde{M}^{-} &\Longrightarrow  &\tilde{M}^{+} \\  
  {\rm Even} & p=2 &\tilde{M}^{++} &\Longrightarrow  &\tilde{M}^{-+}  
  \end{array} \label{eq:NS-point}
 \end{align}
  On the BZ face, a point node in TRS breaking odd-parity SCs is protected by the $\mathbb{Z}_2$ topological number, whereas one in time-reversal invariant even-parity SCs is protected by the $2\mathbb{Z}$ topological number. Similarly to the line node, the nonsymmorphic symmetry plays a crucial role in protecting these point nodes because a point node is topologically unstable in the mirror-invariant plane at $k_{\perp} = 0$.  
 
 \section{Summary} 
 \label{sec:summary}
  We argued the topological stability of nodes in nonsymmorphic SCs, with taking into account the influence of the factor system on the topological classification. The important point is that nonsymmorphic symmetry leads to a nontrivial factor system at BZ faces, which is reflected as the change of the commutation relation between spatial-inversion and mirror-reflection operators in some cases. Adding a nontrivial factor system in the topological classification allows us to deal with a node in nonsymmorphic SCs in the same manner as symmorphic SCs. Although we focused on the order-two symmetries in this paper, Eqs~(\ref{eq:asso2}) and (\ref{eq:Uk0-asso2}) are generally satisfied for all of space group operations, but we need a topological method beyond the Clifford algebra extension method that is outside the scope of this paper. 
  
  In the topological approach, we found nonsymmorphic symmetry protected line (\ref{eq:NS-line}) and point nodes (\ref{eq:NS-point}), which can be considered to be gapless superconducting states analogous to nonsymmorphic symmetry protected topological semimetals. Therefore, our findings will enlarge the category of topological gapless phases and facilitate understanding of gapless superconductors with a nonsymmorphic crystal structure such as UPt$_3$. 
  
  We briefly comment on the bulk-boundary correspondence for nonsymmetry protected line nodes. Usually, a line node induced surface zero-energy flat band is robust as long as the line node is protected by non-spatial symmetry $CT$~\cite{Sato:2011,Schnyder:2015}. In contrast, a crystal symmetry supported surface zero-energy flat band accidentally occurs, so one vanishes by adding crystal symmetry breaking perturbations such as the surface Rashba SOC~\cite{Kobayashi:2015}. In addition, making a surface parallel to a line node may break nonsymmorphic symmetry. For this reason, we expect that the surface flat band induced by a nonsymmetry protected line node may be unstable unless other mechanisms protect it.    
 
 \section{Acknowledgements}
This work was supported in part by the “Topological Materials Science” Grant-in Aid for Scientific Research on Innovative Areas from the MEXT of Japan (No. 15H05855, 16H00991) and a Grant-in-Aid for Scientific Research B (No. 25287085) (MS), and Grant-in-Aid for Scientific Research C (15K05164) (YY). The work of MS was performed in part at the Aspen Center for Physics, which is supported by National Science Foundation grant PHY-1066293.
 
 \appendix
 \section{Group operation on $c_{\bm{k},\alpha}^{\dagger}$}
 \label{app:action}
 Here, we define the group operation on $c_{\bm{k},\alpha}^{\dagger}$. For an element $ \{g| \bm{\tau} \}$ of a space group $G$, the group operation is defined by
 \begin{align}
  &\{ g|\bm{\tau} \} c_{\bm{k},\alpha}^{\dagger} \{ g|\bm{\tau}\}^{-1} \notag \\ 
  &= \frac{1}{\sqrt{N}} \sum_{\bm{R}} e^{i \bm{k}\cdot (\bm{R}+\bm{r}_{\alpha})} c_{\beta}^{\dagger} (D(g)\bm{R} +\bm{\Delta}_{\beta \alpha} +\bm{r}_{\beta} ) U_{\beta \alpha} (g) , \label{eq:wf_sym1}
 \end{align}
with $\bm{\Delta}_{\beta \alpha} = D(g) \bm{r}_{\alpha}-\bm{r}_{\beta} + \bm{\tau}$. If the system is invariant under $G$, there exists a BL vector $\bm{R}'$ such that $\bm{R}' = D(g) \bm{R} + \bm{\Delta}_{\beta \alpha}$ for an arbitrary $\{ g | \bm{\tau}\} \in G$. Then Eq.~(\ref{eq:wf_sym1}) reduces to
 \begin{align}
  \{ g|\bm{\tau} \} c_{\bm{k},\alpha}^{\dagger} \{ g|\bm{\tau}\}^{-1} = e^{- iD(g) \bm{k} \cdot \bm{\tau}} c_{D(g)\bm{k},\beta}^{\dagger} U_{\beta \alpha} (g), \label{eq:wf_sym2}
 \end{align} 
where we use $\bm{k} \cdot \bm{r} = D(g)\bm{k} \cdot D(g)\bm{r}$. A similar definition of the group operation is given in Ref.~\onlinecite{Alexandradinata:2016}. 
 
 \section{Derivation of Eq.~(\ref{eq:Uk0-asso1})}
 \label{app:Uk0}
 The matrix element of the combination of $\tilde{U}_{\bm{k}_0}(g_1)$ and $\tilde{U}_{\bm{k}_0}(g_2)$ is described by
 \begin{align}
  [\tilde{U}_{\bm{k}_0}(g_1) \tilde{U}_{\bm{k}_0}(g_2)]_{\alpha \gamma} = \sum_{\beta} e^{i\theta(g_1,g_2)} \tilde{U}(g_1)_{\alpha \beta} \tilde{U}(g_2)_{\beta \gamma}, \label{eq:Uk0-asso}
 \end{align}
with $\theta(g_1,g_2) = (D(g_1)\bm{k}_0 -\bm{k}_0) \cdot \bm{r}_{\alpha} + (D(g_2)\bm{k}_0 -\bm{k}_0) \cdot \bm{r}_{\beta} $. By the symmetry of the L\"{o}wdin orbitals, if $U(g_1)_{\alpha \beta} \neq 0$, there exists a BL vector $\bm{R}'$ for the inverse element of $\{g_1 | \bm{\tau}_1\}$ such that $D(g_1)^{-1} \bm{r}_{\alpha} -\bm{r}_{\beta} = D(g_1)^{-1} \bm{\tau}_1 + \bm{R}_{g_1}$, where $\bm{R}_{g_1} := \bm{R}' - D(g_1)^{-1} \bm{R}$ is the BL vector. Using this property, $\theta(g_1,g_2)$ is rewritten as
 \begin{widetext}
 \begin{align}
  \theta (g_1,g_2) = \bm{k}_0 \cdot \{  D(g_1)^{-1}\bm{\tau}_1 -D(g_2)^{-1}D(g_1)^{-1} \bm{\tau}_1 + D(g_2)^{-1}D(g_1)^{-1}\bm{r}_{\alpha} -\bm{r}_{\alpha} + \bm{R}_{g_1} -D(g_2)^{-1}\bm{R}_{g_1} \}. \label{eq:phase-g1g2}
 \end{align}
 \end{widetext}
 Substituting Eq.~(\ref{eq:phase-g1g2}) to Eq.~(\ref{eq:Uk0-asso}), we obtain
 \begin{align}
  \tilde{U}_{\bm{k}_0}(g_1) \tilde{U}_{\bm{k}_0}(g_2) = \omega_{g_1,g_2}^{\bm{k}_0} \tilde{U}_{\bm{k}_0}(g_1g_2), \label{eq:AppB}
 \end{align}
 where the third and fourth terms in Eq.~(\ref{eq:phase-g1g2}) becomes
 \begin{align}
  &e^{i\bm{k}_0 \cdot \{D(g_2)^{-1}D(g_1)^{-1}\bm{r}_{\alpha} -\bm{r}_{\alpha}\}} [\tilde{U}(g_1) \tilde{U}(g_2)]_{\alpha \gamma} \notag \\
  &= e^{i\{D(g_1g_2)\bm{k}_{0} -\bm{k}_0\} \cdot \bm{r}_{\alpha} } \tilde{U}(g_1 g_2)_{\alpha \gamma} \notag \\
  &= \tilde{U}_{\bm{k}_0}(g_1g_2)_{\alpha \gamma},
 \end{align}
 and  the fifth and sixth terms in Eq.~(\ref{eq:phase-g1g2}) vanish such that
 \begin{align}
  e^{i \bm{k}_0 (\bm{R}_{g_1} -D(g_2)^{-1}\bm{R}_{g_1})} = e^{-i (D(g_2)\bm{k}_0-\bm{k}_0)\cdot \bm{R}_{g_1}} =1.
 \end{align}

 \section{Vanishing of the mirror topological number in the absence of SOC}
 \label{app:vanish}
 As shown in Sec.~\ref{subsec:Group}, a line node can be unstable in the absence of SOC even when mirror-odd Cooper pairs at the BZ face. Here, we prove this statement from the topological point of view and show that the instability of a line node occurs irrespective of TRS. 
  We start from the condition that $\{P,U(M_{xy})\} = \{C,\tilde{U}(M_{xy})\} =0$ and SOC is absent. The absence of SOC in the normal Hamiltonian allows spin-rotational symmetry, $[e^{i \theta \bm{n} \cdot \bm{\mathcal{S}} }, H(\bm{k})]=0$, where $\bm{\mathcal{S}} = \frac{1}{2}(s_x,s_y,s_z)$ are the generators of spin rotation and $e^{i \theta \bm{n} \cdot \bm{\mathcal{S}} }$ represents the spin rotation about an $\bm{n}$ axis within $0\le \theta <2 \pi$. Without loss of generality, we can choose $\theta=\pi$ and $\bm{n} \parallel \hat{\bm{x}}$. The spin-rotation operator anti-commutes with the mirror-reflection operator, $\{e^{i \pi \mathcal{S}_x},U(M_{xy})\} =0$. Then, the combination of $P$ and $e^{i \pi \mathcal{S}_x}$ satisfies $[Pe^{i \pi \mathcal{S}_x},U(M_{xy})]=0$, leading to $n_{occ,\lambda}(\bm{k}_{\parallel}) = n_{occ,\lambda}(-\bm{k}_{\parallel})$ for any $\bm{k}_{\parallel}$ and $\lambda$, where $\bm{k}_{\parallel}$ is momentum on the mirror-invariant plane. ($n_{occ,\lambda}(\bm{k}_{\parallel})$ is the number of the occupied states with $\lambda$ at $\bm{k}_{\parallel}$.) Since $\{ C, \tilde{U} (M_{xy})\}=0$, PHS leads to $n_{occ,\lambda}(\bm{k}_{\parallel}) = N_{\lambda} - n_{occ,\lambda}(-\bm{k}_{\parallel}) = N_{\lambda} - n_{occ,\lambda}(\bm{k}_{\parallel}) $, resulting in $n_{occ,\lambda}(\bm{k}_{\parallel}) = \frac{N_{\lambda}}{2}$ for any $\bm{k}_{\parallel}$. That is,  $\mathcal{Q}_{\lambda}=0$. The same argument is applicable to time-reversal invariant odd-parity SCs. In this case, an inversion-symmetric Fermi surface with four-fold degeneracy occurs on the mirror-invariant plane. As a result, the presence of SOC is of significant importance in stabilizing a nodal ring.


\begin{thebibliography}{999}
\bibitem{Legget:1975}
A.~J.~Legget, Rev. Mod. Phys. {\bf 47}, 331 (1975).
\bibitem{Sigrist:1991}
M.~Sigrist and K.~Ueda, Rev.\ Mod.\ Phys. {\bf 63}, 239 (1991).
\bibitem{Pfleiderer:2009}
C.~Pfleiderer, Rev.\ Mod.\ Phys.\ {\bf 81}, 1551 (2009).
\bibitem{Anderson:1984}
P.~W.~Anderson, Phys.\ Rev.\ B {\bf 30}, 4000 (1984).
\bibitem{Blount:1985}
E.~I.~Blount, Phys.\ Rev.\ B, {\bf 32}, 2935 (1985). 
\bibitem{Joynt:2002}
R.~Joynt and L.~Taillefer, Rev.\ Mod.\ Phys.\ {\bf 74}, 235 (2002).
\bibitem{Norman:1995}
M.~R.~Norman, Phys. Rev. B, {\bf 52}, 15093 (1995).
\bibitem{Micklitz:2009}
T.~Micklitz and M.~R.~Norman, Phys. Rev. B, {\bf 80}, 100506(R) (2009).
\bibitem{Volovik:2003}
G.~E.~Volovik, {\it The Universe in a Helium Droplet}, (Oxford University Press, New York, 2003).
\bibitem{Sato:2006}
M.~Sato, Phys.\ Rev.\ B, {\bf 73}, 214502 (2006).
\bibitem{Beri:2010}
B.~B\'{e}ri, Phys. Rev. B {\bf 81}, 134515 (2010).
\bibitem{Horava:2005}
P.~Ho\v{r}ava, Phys.\ Rev\ Lett. {\bf 95}, 016405 (2005).
\bibitem{Zhao:2013v1}
Y.~X.~Zhao and Z.~D.~Wang, Phys.\ Rev\ Lett. {\bf 110}, 240404 (2013).
\bibitem{Shiozaki:2014}
K. Shiozaki and M. Sato, Phys.\ Rev.\ B\ {\bf 90}, 165114 (2014).
\bibitem{Kobayashi:2014}
S. Kobayashi, K. Shiozaki, Y. Tanaka, and M. Sato, Phys. Rev. B, {\bf 90}, 024516 (2014).
\bibitem{Chiu:2014}
C.-K.~Chiu and A.~P.~Schnyder, Phys.\ Rev.\ B\ {\bf 90}, 205136 (2014).
\bibitem{SAYang:2014}
S.~A.~Yang, H.~Pan, and F.~Zhang, Phys.\ Rev.\ Lett.\ {\bf 113}, 046401 (2014).
\bibitem{Chiu:2015}
C.-K,~Chiu, J.~C.~Y.~Teo, A.~P.~Schnyder, and S.~Ryu, Rev.\ Mod.\ Phys.\ {\bf 88}, 035005 (2016).
\bibitem{Tanaka:2010}
Y.~Tanaka, Y.~Mizuno, T.~Yokoyama, K.~Yada, and M.~Sato, Phys.\ Rev.\ Lett.\ {\bf 105}, 097002 (2010).
\bibitem{Sato:2011}
M.~Sato, Y.~Tanaka, K.~Yada, and T.~Yokoyama, Phys.\ Rev.\ B {\bf 83}, 224511 (2011).
\bibitem{Yada:2011}
K.~Yada, M.~Sato, Y.~Tanaka, and T.~Yokoyama, Phys. Rev. B {\bf 83}, 064505 (2011).
\bibitem{Tanaka:2012}
Y.~Tanaka, M.~Sato, and N.~Nagaosa, J.\ Phys.\ Soc.\ Jpn. {\bf 81}, 011013 (2012).
\bibitem{Schnyder:2011}
A.~P.~Schnyder and S.~Ryu, Phys.\ Rev.\ B {\bf 84}, 060504(R) (2011).
\bibitem{Brydon:2011}
P.~M.~R.~Brydon, A.~P.~Schnyder, and C.~Timm, Phys. Rev. B {\bf 84}, 020501(R) (2011).
\bibitem{Schnyder:2012}
A.~P.~Schnyder, P.~M.~R.~Brydon, and C.~Timm, Phys.\ Rev.\ B {\bf 85}, 024522 (2012).
\bibitem{Matsuura:2013}
S.~Matsuura, P.-Y~Chang, A.~P.~Schnyder, and S.~Ryu, New J.\ Phys. {\bf 15}, 065001 (2013).
\bibitem{Schnyder:2015}
A.~P.~Schnyder and P.~M.~R.~Brydon, J.\ Phys.; Condens. Matter {\bf 27}, 243201 (2015).
\bibitem{Kobayashi:2015}
S.~Kobayashi, Y.~Tanaka, and M.~Sato, Phys.\ Rev.\ B {\bf 92}, 214514 (2015).
\bibitem{Murakami:2007}
S.~Murakami, New J. Phys. {\bf 9}, 356 (2007).
\bibitem{Young:2012}
S.~M.~Young, S.~Zaheer, J.~C.~Y.~Teo, C.~L.~Kane, E.~J.~Mele, and A.~M.~Rappe, Phys.\ Rev.\ Lett.\ {\bf 108}, 140405 (2012).
\bibitem{Wang:2012}
Z.~Wang, Y.~Sun, X.-Q.~Chen, C.~Franchini, G.~Xu, H.~Weng, X.~Dai, and Z.~Fang, Phys.\ Rev.\ B {\bf 85}, 195320 (2012).
\bibitem{Wang:2013}
Z. Wang, H. Weng, Q. Wu, X. Dai, and Z. Fang, Phys.\ Rev.\ B {\bf 88}, 125427 (2013).
\bibitem{Steinberg:2014}
J.~A.~Steinberg, S.~M.~Young, S.~Zaheer, C.~L.~Kane, E.~J.~Mele, and A.~M.~Rappe, Phys.\ Rev.\ Lett.\ {\bf 112}, 036403 (2014).
\bibitem{Yang:2014}
B.-J.~Yang and N.~Nagaosa, Nat.\ Commun. {\bf 5}, 4898 (2014).

\bibitem{Xu:2014}
S.-Y.~Xu, {\it et al.}, Science {\bf 347}, 294 (2015). 
\bibitem{Liu1:2014}
Z.~K.~Liu, {\it et al.}, Science {\bf 343}, 864 (2014).
\bibitem{Neupane:2014}
M. Neupane, {\it et al.}, Nature Commun. {\bf 5}, 3786 (2014).
\bibitem{Liu2:2014}
Z.~K.~Liu, {\it et al.}, Nature Materials {\bf 13}, 677 (2014).
\bibitem{Jeon:2014}
S.~Jone, {\it et al.}, Nature Materials {\bf 13}, 851 (2014).
\bibitem{Yi:2014}
H.~Yi, {\it et al.}, Sci. Rep. {\bf 4}, 6106 (2014).
\bibitem{Borisenko:2014}
S.~Borisenko, {\it et al.}, Phys. Rev. Lett. {\bf 113}, 027603 (2014). 


\bibitem{Wan:2011}
X.~Wan, A.~M.~Turner, A.~Vishwanath, and S.~Y.~Savrasov, Phys.\ Rev.\ B, {\bf 83}, 205101 (2011).
\bibitem{Burkov:2011}
A.~A.~Burkov, M.~D.~Hook, and L.~Balents, Phys.\ Rev.\ B {\bf 84}, 235126 (2011).
\bibitem{Weng:2015v1}
H.~Weng, C.~Fang, Z.~Fang, B.~A.~Bernevig, and X.~Dai, Phys. Rev. X {\bf 5}, 011029 (2015).
\bibitem{YKim:2015}
Y.~Kim, B.~J.~Wieder, C.~L.~Kane, and A.~M.~Rappe, Phys.\ Rev.\ Lett.\ {\bf 115}, 036806 (2015).
\bibitem{Yu:2015}
R. Yu, H. Weng, Z. Fang, X. Dai, and X. Hu, Phys.\ Rev.\ Lett.\ {\bf 115}, 036807 (2015).
\bibitem{Phillips:2014}
M.~Phillips and V.~Aji, Phys.\ Rev.\ B {\bf 90}, 115111 (2014).
\bibitem{Mullen:2015}
K.~Mullen, B.~Uchoa, and D.~T.~Glatzhofer, Phys.\ Rev.\ Lett.\ {\bf 115}, 026403 (2015).
\bibitem{Xie:2015}
L.~S.~Xie, L.~M.~Schoop, E.~M.~Seibel, Q.~D.~Gibson, W.~Xie, and R.~J.~Cava, APL Mater. {\bf 3}, 083602 (2015).
\bibitem{Weng:2015v2}
H.~Weng, Y.~Liang, Q.~Xu, R.~Yu, Z.~Fang, X.~Dai, and Y.~Kawazoe, Phys.\ Rev.\ B {\bf 92}, 045108 (2015).
\bibitem{Chen:2015v1}
Y.~Chen, Y.~Xie, S.~A.~Yang, H.~Pan, F.~Zhang, M.~L.~Cohen, and S.~Zhang, Nano\ Lett.\ {\bf 15}, 6974 (2015).
\bibitem{Yamakage:2016}
A.~Yamakage, Y.~Yamakawa, Y.~Tanaka, and Y.~Okamoto, J.\ Phys.\ Soc.\ Jpn.\ {\bf 85}, 013708 (2016).
\bibitem{Bian:2016v1}
G.~Bian, {\it et al.} Nat.\ Commun. {\bf 7}, 10556 (2016)..
\bibitem{Bian:2016v2}
G.~Bian, {\it et al.} Phys.\ Rev.\ B {\bf 93}, 121113(R) (2016).
\bibitem{Carter:2012}
J.-M.~Carter, V.~V.~Shankar, M.~A.~Zeb, and H.-Y.~Kee, Phys.\ Rev.\ B {\bf 85}, 115105 (2012).
\bibitem{Chen:2015v2}
Y.~Chen, Y.-M.~Lu, and H.-Y.~Kee, Nat.\ Commun. {\bf 6}, 6593 (2015).
\bibitem{HSKim:2015}
H.-S.~Kim, Y.~Chen, and H.-Y.~Kee, Phys.\ Rev.\ B {\bf 91}, 235103 (2015).
\bibitem{Fang:2015v1}
C.~Fang, Y.~Chen, H.-Y.~Kee, and L.~Fu, Phys.\ Rev.\ B {\bf 92}, 081201(R) (2015).
\bibitem{Watanabe:2015}
H.~Watanabe, H.~C.~Po, A.~Vishwanath, and M.~Zaletel, PNAS {\bf 112}, 47 (2015).
\bibitem{SMYoung:2015}
S.~M.~Young and C.~L.~Kane, Phys.\ Rev.\ Lett.\ {\bf 115}, 126803 (2015).
\bibitem{Watanabe:2016}
H.~Watanabe, H.~C.~Po, M. P.~Zaletel, and A.~Vishwanath, Phys.\ Rev.\ Lett.\ {\bf 117}, 096404 (2016)
\bibitem{Liang:2016}
Q.-F.~Liang, J.~Zhou, R.~Yu, Z.~Wang, and H.~Weng, arXiv:1601.01440.
\bibitem{Wieder:2016}
B.~J.~Wieder and C.~L.~Kane, arXiv:1604.08630.
\bibitem{BJYang:2016}
B.-J.~Yang, T.~A.~Bojesen, T.~Morimoto, and A.~Furusaki, arXiv:1604.00843.
\bibitem{XYZhan:2016}
Y.~X.~Zhan and A.~P.~Schnyder, arXiv:1606.03698.
\bibitem{Mong:2010}
R.~S.~K. ~Mong, A.~M.~~Essin, and J.~E.~Moore, Phys.\ Rev.\ B {\bf 81}, 245209 (2010).
\bibitem{CXLiu:2014}
C.-X.~Liu, R.-X.~Zhang, and B.~K.~VanLeeuwen, Phys.\ Rev.\ B {\bf 90}, 085304 (2014).
\bibitem{Fang:2015v2}
C.~Fang and L.~Fu, Phys. Rev. B {\bf 91}, 161105(R) (2015).
\bibitem{Shiozaki:2015}
K.~Shiozaki, M.~Sato, and K.~Gomi, Phys.\ Rev.\ B {\bf 91}, 155120 (2015). 
\bibitem{Dong:2016}
X.-Y.~Dong and C.-X.~Liu, Phys.\ Rev.\ B {\bf 93}, 045429 (2016).
\bibitem{Varjas:2015}
D.~Varjas, F.~de~Juan, and Y.-M.~Lu, Phys.\ Rev.\ B {\bf 92}, 195116 (2015).
\bibitem{Sahoo:2015}
S.~Sahoo, Z.~Zhang, and J.~C.~Y.~Teo, arXiv:1509.07133.
\bibitem{LLu:2016}
L.~Lu, C.~Fang, L.~Fu, S.~G.~Johnson, J.~D.~Joannopoulos, and M.~Solja\v{c}i\'{c}, Nat.\ Phys.\ {\bf 12}, 337 (2016).
\bibitem{ZWang:2016}
Z.~Wang, A.~Alexandradinata, R.~J.~Cava, and B.~A.~Bernevig, Nature (London) {\bf 532}, 189 (2016).
\bibitem{Alexandradinata:2016}
A.~Alexandradinata, Z.~Wang, and B.~A.~Bernevig, Phys.\ Rev.\ X {\bf 6}, 021008 (2016).
\bibitem{Shiozaki:2016}
K.~Shiozaki, M.~Sato, and K.~Gomi, Phys.~Rev.~B {\bf 93}, Phys.\ Rev.\ B {\bf 93}, 195413 (2016).
\bibitem{QZWang:2016v2}
Q.-Z.~Wang and C.-X.~Liu, Phys. Rev. B {\bf 93}, 020505(R) (2016). 
\bibitem{PYChang:2016}
P.-Y.~Chang, O. Erten, and P. Coleman, arXiv:1603.03435v1. 

 
\bibitem{footnote1}
In this paper, we adopt the notation of crystal symmetries in Ref.~\onlinecite{Alexandradinata:2016}.
\bibitem{Lowdin:1950}
P.-O. L\"{o}wdin, J. Chem. Phys. {\bf 18}, 365 (1950).
\bibitem{Alexandradinata:2014}
A.~Alexandradinata, C.~Fang, M.~J.~Gilbert, and B.~A.~Bernevig, Phys. Rev. Lett. {\bf 113}, 116403 (2014).
\bibitem{Bradley:1972}
C.~J.~Bradley and A.~P.~Cracknell, {\it The Mathematical Theory of Symmetry in Solids} (Clarendon, Oxford, 1972).
\bibitem{Morimoto:2013}
T.~Morimoto and A.~Furusaki, Phys.\ Rev.\ B {\bf 88}, 125129 (2013).
\bibitem{Morimoto:2014}
T.~Morimoto and A.~Furusaki, Phys.\ Rev.\ B {\bf 89}, 235127 (2014).
\bibitem{Yanase:2016}
Y.~Yanase, arXiv:1606.08563.
\bibitem{Fischer:2011}
M.~H.~Fischer, F.~Loder, and M.~Sigrist, Phys.\ Rev.\ B {\bf 84}, 184533 (2011).
\bibitem{Maruyama:2012}
D.~Maruyama, M.~Sigrist, and Y.~Yanase, J. Phys. Soc. Jpn. {\bf 81}, 034702 (2012).
 \end{thebibliography}
\end{document}